\documentclass[aps,showpacs,pre,superscriptaddress,twocolumn]{revtex4-1}
\usepackage{booktabs}
\usepackage{graphicx,amssymb}
\usepackage{mathtools}
\usepackage{physics}
\usepackage{braket}
\usepackage{color}

\begin{document}

\title{Orbital state manipulation of a diamond nitrogen-vacancy center using a mechanical resonator}

\author{H.Y. Chen}
\author{E. R. MacQuarrie}
\author{G. D. Fuchs}
\email{gdf9@cornell.edu}
\affiliation{Cornell University, Ithaca, NY, USA 14853}

\date{\today}

\begin{abstract}

We study the resonant optical transitions of a single nitrogen-vacancy (NV) center that is coherently dressed by a strong mechanical drive. Using a gigahertz-frequency diamond mechanical resonator that is strain-coupled to an NV center's orbital states, we demonstrate coherent Raman sidebands out to the ninth order and orbital-phonon interactions that mix the two excited-state orbital branches.  These interactions are spectroscopically revealed through a multi-phonon Rabi splitting of the orbital branches which scales as a function of resonator driving amplitude, and is successfully reproduced in a quantum model. Finally, we discuss the application of mechanical driving to engineering NV center orbital states.

\end{abstract}
\maketitle

Quantum control of diamond NV centers has enabled scientific accomplishments ranging from fundamental tests of quantum mechanics~\cite{Hensen15} to precision metrology~\cite{Maze08, Barsen17, Dolde11} and quantum information science~\cite{Childress13}. Full state control in the NV center spin manifold has been realized through several techniques developed over the last decade, including magnetic~\cite{Oort88, Jelezko, Fuchs08}, optical~\cite{Golter14, Chu15, Awschalom} and mechanical~\cite{EvanPRL, EvanOptica, JayichNature, PatrickPRL, EvanNature} methods. In contrast, manipulation of an NV center within its excited-state orbital manifold is both less explored and more difficult because direct transitions between excited-state orbital branches through a magnetic dipole interaction are forbidden. However, the intrinsic coupling between NV center orbital states and lattice strain~\cite{DavisHammer, A.Batalov} offers a promising approach to coherently engineering NV center orbital states, potentially enabling quantum information transfer between NV centers, phonon thermometry~\cite{2013PRB}, cooling of a mechanical resonator~\cite{2016PRA}, and decoherence protection of orbital states. 


These applications have drawn growing interest in NV center orbital-strain interactions. For example, wavelength and polarization tuning of NV center optical transitions~\cite{JayichPRA} have been demonstrated using quasi-static lattice strain modulation within a diamond cantilever. Recently, Raman sideband transitions from a strain-modulated NV center have been observed using a surface acoustic wave device and used to manipulate the spin using $\Lambda$-system dynamics near the sideband~\cite{HailinPRL, HailinPRX}. These pioneering experiments motivate additional research aimed at understanding the essential tools and interactions that enable strain-based coherent orbital control of NV centers. Of particular interest are NV-resonator systems that operate in the resolved-sideband limit in which gigahertz-frequency strain oscillations can be treated as coherent phonons.

In this letter we directly engineer the NV center excited orbital manifold using coherent phonons generated by a gigahertz-frequency mechanical resonator. In addition to observing up to nine coherent Raman sideband orders, we demonstrate phonon-induced Rabi splitting of the NV center excited orbital states as a function of resonator driving amplitude and show a signature of multi-phonon orbital transitions. Finally, we show that such orbital-phonon interactions apply to the full spin-orbital fine structure of the NV center excited states.

\begin{figure}
\includegraphics{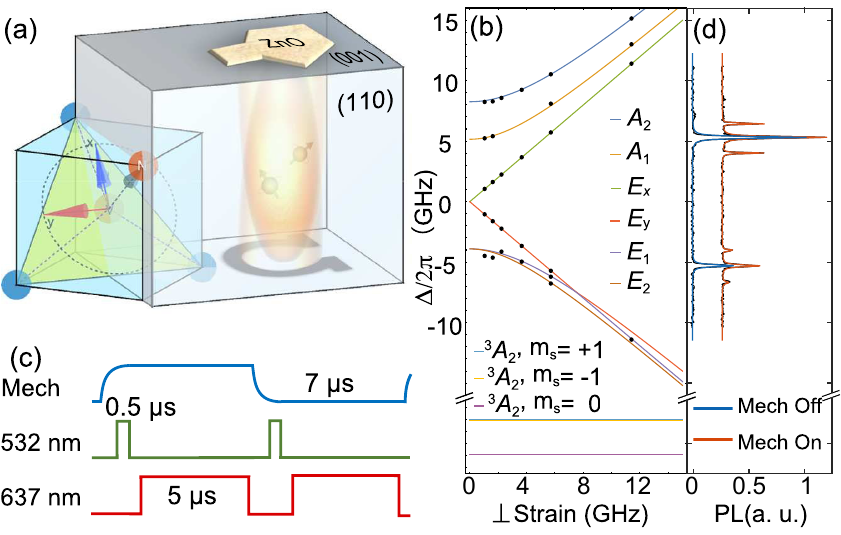}
\caption{(a) Device geometry and structure of an NV center inside a diamond unit cell (blue cube). A uni-axial stress wave is generated through the top zinc oxide transducer. Microwave antenna on the bottom provides magnetic manipulation of NV center spin. (b) PLE sequence for phonon-dressed state measurement. (c) NV center excited-state and ground-state level diagram as a function of transverse strain. Experimental data (black dots) show excellent agreement with the reference~\cite{A.Batalov}. (d) PLE spectrum of a single NV center with 10.6 GHz static strain splitting between $E_{x}$ and $E_{y}$ states (blue curve). A 32 $\mu$W drive to the transducer excites phonon sideband transitions (red curve).}
\label{fig:1}
\end{figure}


For our experiment we fabricated a diamond high-overtone bulk-mode acoustic resonator (HBAR) device that consists of a zinc oxide piezoelectric transducer patterned on the surface of a type IIa diamond substrate, purchased from Element Six [Fig.~\ref{fig:1}(a)]. We designed the thickness of the piezoelectric layer to accommodate a mechanical resonance centered around $\omega_{m}/2\pi$ $\simeq $ 1.3 GHz. On driving the transducer, a longitudinal strain wave enters the acoustic cavity formed by the parallel [100] diamond surfaces. NV centers located at an anti-node of the resulting standing wave experience the strongest strain modulation. The device in this work features resonator modes spanning a range from 1.0 to 1.6 GHz, separated by a free spectral range of 16.7 MHz. On the opposite diamond face we also patterned a microwave antenna for conventional magnetic resonance manipulation of the NV center spin.  Similarly designed devices have been demonstrated to produce intense, high-frequency acoustic waves for strain-based spin manipulation~\cite{EvanPRL, EvanPRB}, but these devices are also suitable for a spectroscopic study of strong phonon driving of NV center orbital states.

The orbital-singlet, spin-triplet ground states of an NV center, $\ket{^{3}A_{2}}$, are coupled to the first excited-state manifold, $\ket{^{3}E}$, through an optical transition with a zero-phonon line (ZPL) at 637.2 nm. The $\ket{^{3}E}$ manifold consists of six levels, $\ket{E_{x,y}}$, $\ket{E_{1,2}}$ and $\ket{A_{1,2}}$. Strain distortions of the NV centers couple strongly to the the orbital components of $\ket{^3E}$, both globally shifting all the states and splitting them into two orbital branches [Fig.~\ref{fig:1}(b)]~\cite{JahnTeller}. Strain of the NV center can originate either from static stress that is intrinsic to the crystal or from external stress modulation. Using group theory analysis, we can factor the strain coupling into three terms according to the symmetry of the deformation~\cite{Hughs, DresselhausCH3}, $A_{1}$, $E_{1}$ and $E_{2}$. $A_{1}$ induces a global shift of the excited state energies of NV center since it retains the $C_{3v}$ symmetry of the defect. In contrast, $E_{1}$ and $E_{2}$ lower the symmetry and thus split and mix the $E$ states. In the basis of $\ket{E_{x}}$ and $\ket{E_{y}}$ states, the strain Hamiltonian takes the Jahn-Teller form~\cite{book}
 \begin{equation}
\label{eq:strain-Ham}
H=V_{A_{1}}+V_{E_{1}}\sigma_{z}+V_{E_{2}}\sigma_{x}, 
\end{equation}
where $V_{i}$ is the deformation potential of $i$-type symmetry, and $\sigma_{z}$, $\sigma_{x}$ are Pauli matrices.

Although static local strain sets the equilibrium orbital state energies of a particular NV center, coherent phonon perturbations can further modify the orbital structure into phonon-dressed states, $\ket{E_{x,y},n}$, where $n$ is the coherent phonon number. In particular, the uniaxial stress wave generated within the HBAR device excites $A_{1}$ and $E_{1}$ phonon modes~\cite{SupplementaryInfo}. Therefore, the orbital state strain Hamiltonian becomes
 \begin{equation}
\label{eq:strain-Ham2}
H=V_{A_{1}}+\mathcal{A}\text{cos}\omega_{m}t+(V_{E_{1}}+\mathcal{E}_{1}\text{cos}\omega_{m}t)\sigma_{z}+V_{E_{2}}\sigma_{x}, 
\end{equation}
where $\mathcal{A}$ ($\mathcal{E}_{1}$) characterizes the strength of the $A_{1}$ ($E_{1}$) components of the phonons excited at the mechanical resonator frequency $\omega_{m}$.  We retain $V_{A_{1}}$, $V_{E_{1}}$ and $V_{E_{2}}$ from Eq.~(\ref{eq:strain-Ham}) to parametrize the static-strain induced energy shifting and splitting of the NV center.

\begin{figure}
\includegraphics{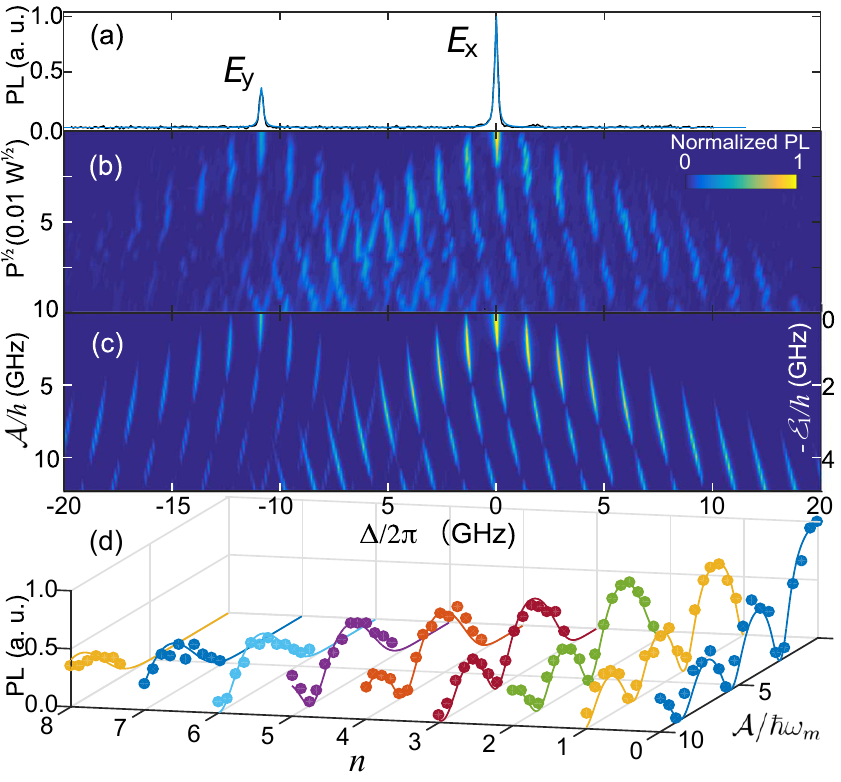}
\caption{(a) PLE spectrum of NV1 with $\{\ket{E_{x}}$, $\ket{E_{y}}\}$ state splitting of 10.6 GHz in the absence of mechanical driving. (b) Phonon-dressed state PLE measurement with $\omega_{m}/2\pi$=1.3844 GHz. The mechanical driving amplitude is proportional to square root of the power applied to the transducer, $P^{1/2}$. Sideband transitions and level repulsion are evident in between $E$ states. (c) Reconstruction of experimental data through quantum master equation simulation. (d) $E_{x}$ sideband transition (n=0 to 8) photoluminescence peak amplitude (solid dots) plotted against theoretical prediction(solid line), which is not a fit to the data.}
\label{fig:2}
\end{figure}

Experimentally, we probe the orbital-phonon dressed states through photoluminescence excitation (PLE) spectroscopy at zero applied magnetic field. The HBAR device is first cooled to 10 K in a helium-flow microscopy cryostat. Using our home-built scanning confocal microscope, we locate and isolate a single NV center. To perform PLE, we sweep the frequency of a 637.2 nm tunable laser across the resonant optical transitions and monitor the incoherent phonon sideband fluorescence above 670 nm with a single photon detector. We use a scanning Fabry-Perot cavity to quantify the tunable laser frequency. Figure \ref{fig:1}(c) shows the experimental sequence that we use to study phonon-dressed optical transitions. First, we turn on the strain modulation for a period of 7 $\mu$s. We include a 2 $\mu$s delay to allow the resonator to ring up (quality factor $\simeq$ 1500). Within that initial delay, we use a 532 nm laser to optically pump the charge of the NV center into the NV$^{-}$ state and the spin into the $m_{s}=0$ state. After ring-up, we turn-on the tunable laser and collect photons for the remaining 5 $\mu$s. As a control, we repeat the cycle except without driving the mechanical resonator.

Because the initialized, unexcited state of the NV center is $\ket{^{3}A_{2}, m_{s}=0}$, resonant optical excitations connect $\ket{^{3}A_{2}, m_{s}=0}$ to $\ket{E_{x}}$ and $\ket{E_{y}}$ in the excited state. Thus we initially restrict our study to pure orbital-strain interactions in the spin-zero-state manifold. Taking into account the photon driving field from the tunable laser, the Hamiltonian of the driven system after applying the rotating wave approximation can be written as
\begin{equation}
\label{eq:FullHam}
\begin{split}
H(m_{s}=0)=&(-\hbar\Delta+\mathcal{A}\text{cos}\omega_{m}t)(\ket{x}\bra{x}+\ket{y}\bra{y})
\\&+(V_{E_{1}}+\mathcal{E}_{1}\text{cos}\omega_{m}t)(\ket{x}\bra{x}-\ket{y}\bra{y})
\\&+V_{E_{2}}(\ket{x}\bra{y}+\ket{y}\bra{x})
\\&+\frac{\hbar}{2}\Omega(\ket{g}\bra{x} +\ket{x}\bra{g}+\ket{g}\bra{y} +\ket{y}\bra{g}),
\end{split}
\end{equation}
where $\ket{x}$, $\ket{y}$ and $\ket{g}$ are simplified notation for $\ket{E_{x}}$, $\ket{E_{y}}$ and the ground state $\ket{^{3}A_{2}, m_{s}=0}$. $\Delta=\omega_{l}-\omega_{0}-V_{A_{1}}/\hbar$ is the laser detuning, $\omega_{0}$ is the unperturbed NV center transition frequency, and $\Omega$ is the optical Rabi frequency.  Due to the inhomogeneity of the stress within the sample, we observed NV centers with $\{\ket{E_{x}}$, $\ket{E_{y}}\}$ state splitting ranging from 2 to 30 GHz [Fig.~\ref{fig:1}(b)]. Fig.~\ref{fig:1}(d) shows PLE spectra taken from an NV center with a static orbital splitting of 10.6 GHz, both with and without (32 $\mu$W to the transducer) mechanical driving.  We immediately see that mechanical driving introduces resolved sidebands to each of the two optical transitions in the PLE spectrum. We also note that because NV centers positioned near an antinode of the resonator are deep within the diamond bulk ($>30~\mu m$ below the diamond surface for the NV centers studied in this paper), charge fluctuations are smaller than for near-surface NV centers~\cite{nl2012}. Thus, we obtain typical PLE linewidths of 100 MHz for optical power of 830 nW. In this work we select NV centers with orbital strain splittings of 10.6, 3.2, and 2.1 GHz, and study their PLE spectra as we vary the driving power to the transducer from 0 to 10 mW. Based on the variation of the PLE linewidth, we estimate that sample heating is less than 5 K.


First we study an NV center (labeled NV1) with a static strain splitting of 10.6 GHz [Fig.~\ref{fig:2}(a)], which is many times larger than $\omega_m/2\pi=$~1.3844 GHz. In the presence of mechanical driving, coherent Raman sideband transitions appear at $\Delta=\pm n\omega_{m}$ with respect to the undriven optical resonance frequencies due to photon-phonon coupling. Specifically, it arises from $A_{1}$ modulation of the $E$ states through the $\mathcal{A}\text{cos}\omega_{m}t$ term in Eq.~(\ref{eq:FullHam})  [Fig.~\ref{fig:2}(b)].  This effect is from sideband-resolved frequency modulation, which can arise in both classical and quantum resonances~\cite{PatrickPRL, microwave, QD, Rydberg,reviewModernPhysics}.

As the mechanical driving strength increases, up to $n=9$ sideband orders emerge at a driving power of 10 mW [Fig.~\ref{fig:2}(b)]. We expect that the amplitude of each sideband transition varies according to a Bessel function form, $J^{2}_{n}(\mathcal{A}/\hbar\omega_{m})$, which can be derived from either the rotating wave approximation~\cite{SCQ} or Floquet theory~\cite{theorypaperSQC}. Because of laser-power broadening, however, the sideband order peaks are described by $\frac{s_{0}J^{2}_{n}(\mathcal{A}/\hbar\omega_{m})}{1+s_{0}J^{2}_{n}(\mathcal{A}/\hbar\omega_{m})}$ [Fig.~\ref{fig:2}(d)], where $s_{0}=2\Omega ^{2}/\Gamma ^{2}$ is the saturation parameter and $\Gamma$ is the linewidth of NV center optical transition~\cite{MetCalfbook}.

A second prominent feature in the dressed spectral evolution of NV1 is the repulsive frequency shift of the two orbital branches, including sidebands, as a function of mechanical driving amplitude: $\ket{E_{y}}$ ($\ket{E_{x}}$) branches are subject to red (blue) shifts in frequency [Fig.~\ref{fig:2}(b)]. 
This is caused by $E_{1}$-modulation-induced Rabi splitting of the $E$ states in the presence of non-zero $V_{E_{2}}$, which becomes clear by applying a polaron transformation~\cite{JCP, PRA2007, 2010PRB} to the orbital state strain Hamiltonian. In the displaced oscillator orbital basis, Eq.~(\ref{eq:strain-Ham2}) becomes~\cite{SupplementaryInfo}
\begin{widetext}
\begin{equation}
\label{eq:RWT2}
\begin{split}
{H}'&=
\begin{pmatrix}
-\hbar\omega_{m}+\Delta_{x}+\mathcal{A}\text{cos}\omega_{m}t  & -\mathcal{E}_{1}\text{sin}2\theta\sum\limits_{n}  J_{n}(2\frac{\mathcal{E}_{1}}{\hbar\omega_{m}}\text{cos}2\theta)e^{-in\omega_{m}t}\\ 
-\mathcal{E}_{1}\text{sin}2\theta\sum\limits_{n}  J_{n}(2\frac{\mathcal{E}_{1}}{\hbar\omega_{m}}\text{cos}2\theta)e^{in\omega_{m}t} & -\Delta_{x}+\mathcal{A}\text{cos}\omega_{m}t 
\end{pmatrix}
\end{split}
\end{equation}
\end{widetext}
where $\theta$ is the mixing angle of the static strain deformation potential defined by tan$2\theta=V_{E_{2}}/V_{E_{1}}$; $2\Delta_{x}=2\sqrt{V^{2}_{E_{1}}+V^{2}_{E_{2}}}$ is the static strain splitting of $E$ states, and $J_{n}$ is the Bessel function of the first kind of order $n$. The off-diagonal terms in Eq.~(\ref{eq:RWT2}) characterize the ($n$+1)-phonon coupling of orbital states from $E_{1}$ modulation, which give rise to multi-phonon driven orbital transitions~\cite{SupplementaryInfo} when the resonance condition $(n+1)\hbar\omega_{m}=2\Delta_{x}$ is satisfied. 

\begin{figure}
\includegraphics{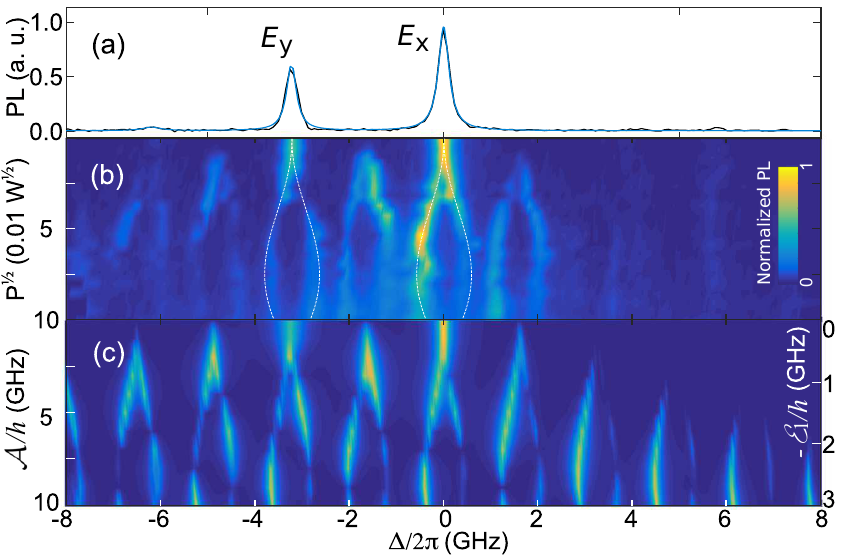}
\caption{(a) PLE scan of NV2, which has an $\{\ket{E_{x}}$, $\ket{E_{y}}\}$ state splitting of 3.2 GHz. (b) Experimental data from phonon-dressed state PLE. The mechanical driving frequency, $\omega_{m}/2\pi$=1.6 GHz, matches half of the $\{\ket{E_{x}}$, $\ket{E_{y}}\}$ splitting. The dressed orbital states exhibit gigahertz-scale phonon Rabi splitting, which is well described by the $n=1$ term in Eq.~(\ref{eq:RWT2}), $\mathcal{E}_{1}\text{sin}2\theta J_{1}(2\frac{\mathcal{E}_{1}}{\hbar\omega_{m}}\text{cos}2\theta)$, which corresponds to a 2-phonon Rabi splitting (white dashed lines). (c) Reconstruction of experimental data with a quantum master equation simulation.}
\label{fig:4}
\end{figure}

To further study this effect, we select a different NV center (NV2) with a comparably small static $E$ state splitting, 3.2 GHz, which we can bring into a 2-phonon resonance by driving with a mechanical mode frequency of $\omega_{m}/2\pi=1.6$ GHz. Figure \ref{fig:4} shows the resulting spectra as a function of $\mathcal{E}_{1}$ amplitude. As $\mathcal{E}_{1}$ increases, the individual orbital levels first split linearly, while the splitting starts to decrease at a higher $\mathcal{E}_{1}$ amplitude. This phonon Rabi splitting is well described by the $n=1$ coupling term in Eq.~(\ref{eq:RWT2}), $\mathcal{E}_{1}\text{sin}2\theta J_{1}(2\frac{\mathcal{E}_{1}}{\hbar\omega_{m}}\text{cos}2\theta)$ [outlined by the white dashed lines in Fig.~\ref{fig:4}(b)], with a small deviation caused by the contribution of off-resonant phonon interactions of other orders, $n+1\neq2$, indicating that the resonant 2-phonon process dominates the dressed orbital states under these conditions.

To quantitatively model the general case of phonon-dressed orbital states, we implement a quantum master equation simulation~\cite{SupplementaryInfo}. We adjust $\mathcal{A}$ and $\mathcal{E}_{1}$ as free parameters by hand to match the calculated spectra to the experiment. The simulations [Fig.~\ref{fig:2}(c), Fig.~\ref{fig:4}(c)] show excellent agreement to the measured spectra. For the case of NV2 with resonant-phonon driving, the gigahertz orbital Rabi splitting [Fig.~\ref{fig:4}(b)] implies that, after laser excitation to one of the orbital states, coherent Rabi oscillations between the two NV center orbital states will occur on the time-scale of a nanosecond (see Supplementary Information~\cite{SupplementaryInfo}), which is an order of magnitude smaller than the intrinsic excited state life time of NV centers~\cite{Batalov08}.

\begin{figure}
\includegraphics{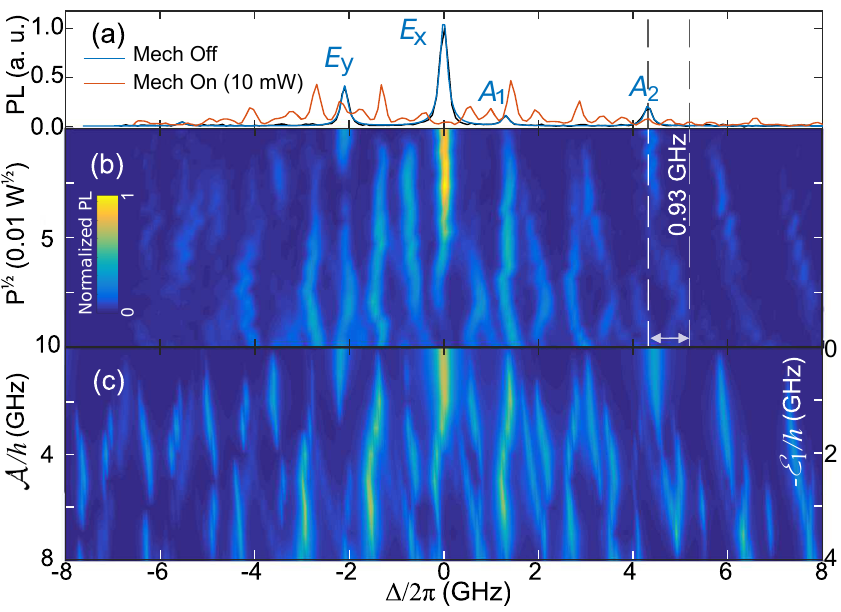}
\caption{(a) PLE scan of NV3 with $\{\ket{E_{x}}$, $\ket{E_{y}}\}$ state splitting of 2.1 GHz, taken with continuous wave 2.877 GHz magnetic microwave driving to the antenna. (b) Experimental data from phonon-dressed state PLE measurement of NV3 with $\omega_{m}/2\pi$=1.3844 GHz. A Rabi splitting of 0.93 GHz in the $\ket{A_{2}}$ state is observed with 10 mW mechanical driving. (c) Reconstruction of experimental data using a quantum master equation simulation. }
\label{fig:3}
\end{figure}

Next, we probe electron-phonon coupling in the full six-level NV center excited states on a third NV center (NV3) with a static strain splitting of 2.1 GHz. To spectroscopically access the orbital states with $m_{s}\neq0$ character, we drive a continuous wave, 2.877 GHz current into the microwave antenna, which prepares NV3 in a mixed spin state prior to resonant laser excitation. This enables access to all excited-state optical transitions, i.e. $\ket{^{3}A_{2}, \abs{m_{s}}=1}\rightarrow\ket{A_{1}}$, $\ket{^{3}A_{2}, \abs{m_{s}}=1}\rightarrow\ket{A_{2}}$, $\ket{^{3}A_{2}, \abs{m_{s}}=1}\rightarrow\ket{E_{1}}$ and $\ket{^{3}A_{2}, \abs{m_{s}}=1}\rightarrow\ket{E_{2}}$ can all be excited, in addition to the ones studied above. A single PLE scan with magnetic microwave driving is shown in Fig.~\ref{fig:3}(a). $\ket{A_{1}}$ and $\ket{A_{2}}$ orbital states are revealed on the right of $E$-state peaks, and $\ket{E_{1,2}}$ are out of the scanning range. The small peak on the left of Fig.~\ref{fig:3}(a) is from a non spin-conserving transition between $\ket{^{3}A_{2}, m_{s}=0}$ and the $\ket{E_{1,2}}$ states. The corresponding phonon dressed state spectra ($\omega_{m}/2\pi=1.3844$ GHz) is shown in Fig.~\ref{fig:3}(b). $\ket{A_{1}}$ and $\ket{A_{2}}$ experience comparable modulation with $\ket{E_{x}}$ and $\ket{E_{y}}$ in response to a coherent phonon driving. Apart from the emergence of sideband transitions, 10 mW mechanical driving produces a phonon Rabi splitting of 0.93 GHz as a result of $E_{1}$-modulation induced mixing between $\ket{A_{1}}$ ($\ket{A_{2}}$) and $\ket{E_{1}}$ ($\ket{E_{2}}$) states [marked by the white arrow in Fig.~\ref{fig:3}(b)]. A resonant phonon field can thus drive orbital transitions between the $m_{s}\neq0$ orbital levels. The dressed state spectra generally agree with our 8-level master equations simulation~\cite{SupplementaryInfo} result in Fig.~\ref{fig:3}(c). We attribute the mismatch in the simulated intensity of $E_{y}$ sidebands to the inter-system crossing process through the singlet states~\cite{ISC}.

We can evaluate the strength of $A_{1}$ and $E_{1}$ modulation generated by the HBAR device for each of the three NV centers by comparing the data and the simulation results. We find $\mathcal{E}_{1}/\mathcal{A}\sim-0.4(1)$ for all three NV centers. At modest mechanical driving power (10 mW), we observe phonon coupling up to $\mathcal{A}/h\simeq13$ GHz. Using the previously measured coupling factor, $465$ Hz/Pa~\cite{DavisHammer, DohertyExp, JayichPRA}, we estimate that the stress amplitude produced by the  resonator is around 30 MPa, corresponding to a lattice strain of $7\times 10^{-6}$, which is consistent with our previous work~\cite{EvanPRL} , and it is comparable to the strain generated in an oscillating cantilever system~\cite{JayichPRA}. As such, dressed state spectroscopy provides a quantitative approach for  AC strain characterization within a resonator (see Supplementary Information~\cite{SupplementaryInfo}).

Phonon-driven orbital transitions also provide a powerful method to control an NV center orbital quantum state dynamically, for example, resonant Rabi driving between $\ket{E_{x}}$ and $\ket{E_{y}}$. Because electric fields and strains shift NV center optical transitions in similar ways~\cite{Maze2011, Doherty2011}, a potential application of orbital Rabi driving is continuous dynamical decoupling (CDD)~\cite{Fanchini07} of orbital states. Orbital CDD would operate in an analogous fashion to CDD of spin states~\cite{Rabl09, Xu12, Golter142, PatrickNature, EvanPRB} : when driving an orbital transition on resonance, the dressed orbital eigenstates have $d\omega/d\epsilon_{\perp}=0$, where $\omega$ is the optical transition frequency, and $\epsilon_{\perp}$ is the electric field strength transverse with respect to the NV axis. This would eliminate the NV center's sensitivity to the transverse components of the uncontrolled, environmental electric field fluctuations originating from, for example, charge repumping or surface charge fluctuations. This will make the optical transitions more robust to spectral diffusion and reduce inhomogeneous line broadening~\cite{spectralDiff}. Therefore, coherent phonon-driven orbital control is an untapped resource for improving the optical properties of NV centers in support of quantum optical control~\cite{Golter14, Chu15, Awschalom}, spin-photon entanglement~\cite{Togan10, Bernien13}, coupling to optical cavities~\cite{Albrecht13}, and other applications of resonant optical coupling.


In summary, we have experimentally examined the resonant optical transitions of an NV center subject to a coherent phonon drive. In the phonon-dressed orbital states, we observe a strong orbital-phonon interaction that gives rise to multiple Raman sidebands and a large phonon-induced Rabi splitting of the orbital states, which are due to $A_{1}$- and $E_{1}$-type phonon modulation, respectively. Finally, we show that resonant mechanical driving leads to multi-phonon transitions between orbital states and discuss a potential application for orbital state decoherence protection.

\begin{acknowledgments}
Research support was provided by the Office of Naval Research (Grant N000141410812 and N000141712290). Device fabrication was performed in part at the Cornell NanoScale Science and Technology Facility, a member of the National Nanotechnology Coordinated Infrastructure, which is supported by the National Science Foundation (Grant ECCS-15420819), and at the Cornell Center for Materials Research Shared Facilities which are supported through the NSF MRSEC program (Grant DMR-1719875). 
\end{acknowledgments}

\clearpage
\onecolumngrid

\centerline{{\huge Supplementary Materials}}

\section{NV center electron orbital structure}

We first start with a short discussion of the equilibrium NV center electron configurations without lattice distortion. One can find more detailed explanations in references ~\cite{Maze2011,Doherty2011}.

\begin{figure}[!ht]
\includegraphics{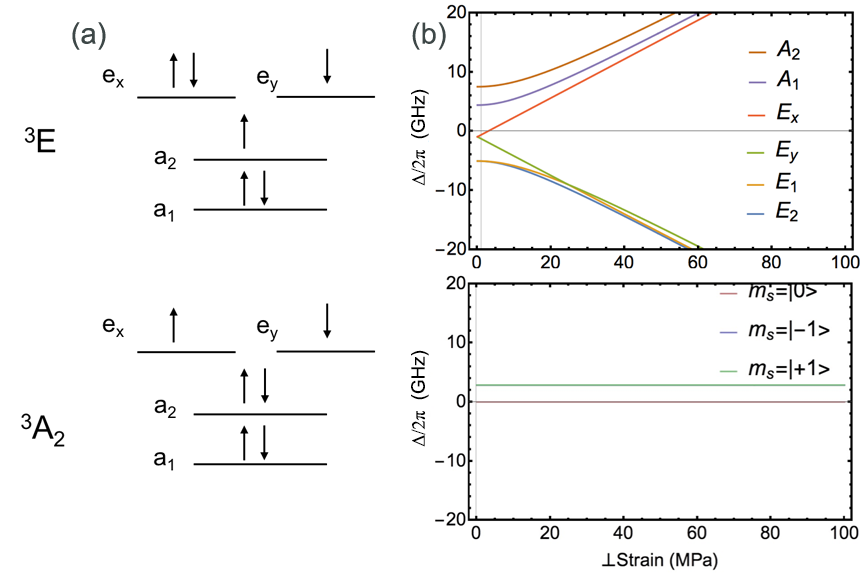}
\caption{(a) NV center electron orbital configuration in the ground and excited state. (b) Strain induced splitting in NV center levels.}
\label{fig:level}
\end{figure}

In diamond, the presence of a vacancy and a neighboring substitution nitrogen atom produces four dangling bonds {$c1,c2,c3,n$} formed from the unpaired electrons. Projecting the character representation of the four bonds onto the irreducible representations of $C_{3v}$ point group symmetry yields four electron orbitals, constructed from linear combination of the bonds: $a_{1}=n, a_{2}=(c1+c2+c3)/3, e_{x}=(2a_{1}-a_{2}-a_{3})/\sqrt{6}$ and $e_{y}=\sqrt{2}(a_{2}-a_{3})$. $a_{1}, a_{2}$ transforms with $A_{1}$ symmetry, and  $e_{x}, e_{y}$ transform with $E$ symmetry. The coulomb interaction mixes $a_{1}$ and $a_{2}$ and enforces the energy ordering of orbitals illustrated in Fig.~\ref{fig:level}(a). 

The orbital configuration of the six electrons from a negatively charged NV center gives the final electronic structure, the basis functions of which determines the electronic wave function. In the ground state, $a_{1}$ and $a_{2}$ are fully filled while there is one electron on each $e$ orbital. This leads to a symmetrized orbital and a spin triplet electron state $\ket{^{3}A_{2}}$. Promoting one electron from $a$ to $e$ produces the optically active excited state $\ket{^{3}E}$, the orbital part wave function of which displays $E$ symmetry [Fig.~\ref{fig:level}(a)]. 

\section{Group theory approach of electron-phonon coupling in the excited $E$ states}

We now focus on the zero spin orbital states of the excited state, $\ket{E_{x}}$ and $\ket{E_{y}}$. As we see in the above discussion, $\ket{E_{x,y}}$ states are degenerate in energy in the equilibrium atomic configuration $Q_{0}$, which displays $C_{3v}$ symmetry. Considering a perturbation $\sum_{r}Q_{r}$ from a combination of linear nuclear displacements drawn from orthogonal set $\hat{n}_{r}$, where $Q_{r}=q_{r}\hat{n}_{r}$ and  $Q_{r}$ transform according to irreducible representation of $r$ group, the new nuclear configuration then reads $Q=Q_{0}+\sum_{r}Q_{r}$. To first order, the Hamiltonian of the NV center electronic wave function can then be expanded as
\begin{equation}
\label{eq:JT0}
H_{e}=H_{0}+V+O(Q^{2}_{r}),
\end{equation}
where $V=\sum_{r}^{ }V_{r} Q_{r}$ is the first order perturbation potential and $V_{r}=\left ( \frac{\partial H_{e}}{\partial Q_{r}}\right )_{0}$ is dependent on the structure of the NV center and can be measured experimentally. $V_{r}$ and $Q_{r}$ have the same transformation properties according to $r$ group symmetry. The eigenenergy of the perturbed system can then be calculated from the Jahn-Teller matrix~\cite{JahnTeller}. In the basis of $\ket{E_{x}}$ and $\ket{E_{y}}$, we have

\begin{equation}
\label{eq:JT}
H_{e}=H_{0}+\begin{pmatrix}
\bra{E_{x}}V \ket{E_{x}}&  \bra{E_{x}}V \ket{E_{y}}\\
 \bra{E_{y}}V \ket{E_{x}} &  \bra{E_{y}}V \ket{E_{y}}
\end{pmatrix},
\end{equation}
where 
\begin{equation}
\label{eq:JT2}
\bra{E_{i}}V \ket{E_{j}}
=\sum_{r}^{ }Q_{r} \bra{E_{i}}V_{r} \ket{E_{j}}
=\sum_{r}^{ }Q_{r} \int \psi^{*}_{E_{i}} \psi_{E_{j}}V_{r} \text{d}\tau, 
\end{equation}
$\psi^{*}(E_{i})\psi(E_{j})$ forms a reducible product representation $E\times E$ which can be further factorized into $E\times E=A_{1}+A_{2}+E$, with the associated basis functions~\cite{book}:

\begin{equation}
\label{eq:JTf}
\begin{split}
f_{A1}&=\frac{1}{2}(\abs{\psi_{E_{x}}}^{2}+\abs{\psi_{E_{y}}}^{2}),
\\
f_{A2}&=\frac{1}{2}(\psi^{*}_{E_{x}}\psi_{E_{y}}-\psi^{*}_{E_{y}}\psi_{E_{x}}),
\\
f_{E_{1}}&=\frac{1}{2}(\abs{\psi_{E_{x}}}^{2}-\abs{\psi_{E_{y}}}^{2}),
\\
f_{E_{2}}&=\frac{1}{2}(\psi^{*}_{E_{x}}\psi_{E_{y}}+\psi^{*}_{E_{y}}\psi_{E_{x}}).
\end{split}
\end{equation}
Note that $V_{r}$ is of symmetry group $r$, one can then apply the orthogonality theorem~\cite{DresselhausCH3} and vanishing integral rule to calculate the integral
\begin{equation}
\label{eq:JTf2}
I_{{r}'}=\int f_{{r}'} V_{r} \text{d}\tau =\delta_{r,{r}'}\int f_{{r}'} V_{{r}'} \text{d}\tau.
\end{equation}
The integral of Eq.~(\ref{eq:JT2}) is non-vanishing only if $r$ is of $A_{1}, A_{2}$ or $E$ symmetry, namely,

\begin{equation}
\label{eq:JTf3}
H_{e}=H_{0}+
\begin{pmatrix}
 I_{A1}Q_{A1}+I_{E_{1}}Q_{E_{1}} & I_{A_{2}}Q_{A_{2}}+I_{E_{2}}Q_{E_{2}}\\ 
- I_{A_{2}}Q_{A_{2}}+I_{E_{2}}Q_{E_{2}} &  I_{A1}Q_{A1}-I_{E_{1}}Q_{E_{1}}
\end{pmatrix}
\end{equation}

The physical interpretation of Eq.~(\ref{eq:JTf3}) is that $\ket{E}$ states will only couple to a perturbation potential from strain or lattice phonons of $A$ or $E$ symmetry, otherwise the energy of the NV center electronic structure will remain unchanged. In particular, an $A_{1}$ perturbation will shift the energy of the $\ket{E}$ state while the overall NV center $C_{3v}$ symmetry is preserved; $E_{1}$ perturbations lie diagonal in Eq.~(\ref{eq:JTf3}) and will symmetrically split the $\ket{E_{x}}$ and $\ket{E_{y}}$ state in energy. $E_{2}$ perturbations are purely off diagonal and thus lead to mixing of the $\ket{E_{x}}$ and $\ket{E_{y}}$ state as well as a splitting in their energies [Fig.~\ref{fig:level}(b)]. $A_{2}$ stands for a rotational symmetry interaction, which is absent for the case of uniaxial strain deformation.

\section{[001] Uniaxial stress on diamond, static and modulation}

In the current experiment, the uni-axial stress perturbation along the [001] direction of the diamond crystal, $\sigma_{ZZ}$, contains $A_{1}-$ and $E$-type of deformation in the cubic representation of $O_{h=6}$ point group. Assuming the atomic displacement induced from stress is linear (diamond has a small Poisson ratio $\sim$0.11), the exact form of the perturbations $Q_{r}$ and $V$ can then be obtained by reducing the $O_{h=6}$ representation to $C_{3v}$. We end up with $A_{1}$ and $E$ perturbations, and the Jahn-Teller matrix then can be constructed using the $C_{3v}$ basis functions as~\cite{Hughs}:

\begin{equation}
\label{eq:JT4}
\begin{split}
V&=V_{A1}(\sigma_{XX}+\sigma_{YY}+\sigma_{ZZ})+V_{E_{1}}(2\sigma_{ZZ}-\sigma_{XX}-\sigma_{YY})+V_{E_{2}}\sqrt{3}(\sigma_{XX}-\sigma_{YY}),
\\
H_{e}&=H_{0}+
\begin{pmatrix}
 (A+2B)(\sigma_{XX}+\sigma_{YY}+\sigma_{ZZ}) &\sqrt{3}B(\sigma_{XX}-\sigma_{YY})\\ 
\sqrt{3}B(\sigma_{XX}-\sigma_{YY}) &  (A-2B)(\sigma_{XX}+\sigma_{YY}+\sigma_{ZZ}),
\end{pmatrix}
\end{split}
\end{equation}
where $X, Y, Z$ are along lattice directions [100], [010] and [001], respectively. $\sigma_{XX}$, $\sigma_{YY}$ and $\sigma_{ZZ}$ are stress tensor elements. $A$ and $B$ are the stress coupling factor. $A$ can be treated as the counterpart of $I_{A1}$ in Eq.~(\ref{eq:JTf}) with the replacement that $V_{A1}=\left ( \frac{\partial H_{e}}{\partial \sigma_{A1}}\right )_{0}$. $B$ can be defined in a similar way for E type perturbation. They have been experimentally measured as $A=1.92\times 10^{-12} \text{eV Pa}^{-1}$ and $B=1.36\times 10^{-12} \text{eV Pa}^{-1}$~\cite{DavisHammer, JayichPRA} (A correction has been taken into account from the measurement in~\cite{DohertyExp}). 

Comparing Eq.~(\ref{eq:JT4}) to Eq.~(\ref{eq:JTf3}), it is readily seen that perfectly aligned static stress along $Z$ ([001]) will induce $A_{1}$ and $E_{1}$ types of interaction. When the applied stress is off axis such that $\sigma_{XX}\neq\sigma_{YY}$, the diagonal term in the JT matrix will remain a finite value and induce mixing of the $E$ states. 

In our experiment, $\sigma_{ZZ}(t)=\sigma_{0}\text{cos}(\omega_{m}t)$ is time varying of modulation frequency $\omega_{m}$. Taking into account the intrinsic local static strain perturbation to the NV center, $H_{\text{int}}$, we can construct the total strain Hamiltonian as

\begin{equation}
\label{eq:JT5}
\begin{split}
H_{e}&=H_{0}+H_{\text{int}}+H_{\sigma_{ZZ}}(t)
\\
H_{\text{int}}&=V_{A_{1}}+V_{E_{1}}\sigma_{z}+V_{E_{2}}\sigma_{x}
\\
H_{\sigma_{ZZ}}(t)&=(A+2B\sigma_{z})\sigma_{0}\text{cos}(\omega_{m} t),
\end{split}
\end{equation}
where $\sigma_{x}$ and $\sigma_{z}$ are Pauli matrices in the basis of $\{\ket{E_{x}},\ket{E_{y}}\}$.

\section{Extraction of intrinsic strain}\label{intrinsic}

 \begin{figure}[!htbp]
 \centering
\includegraphics{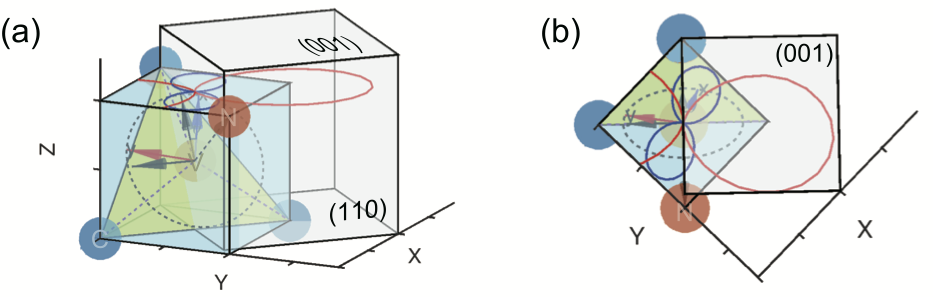}
\caption{(a) Schematic illustration of the dipole moment rotation of an [111] oriented NV center. The transparent cube represents the geometry of the actual diamond crystal sample and the blue cube is the diamond unit cell. Blue and red curves indicate the $\ket{E_{x}}, \ket{E_{y}}$ dipole projection to the (001) surface in response to a linearly polarized laser field incident along [001], plotted as NV photoluminescence (PL) (radial axis) versus the laser polarization angle $\phi$.  (b) (001) view of the same figure in (a).}
\label{fig:crystal}
\end{figure}

The intrinsic strain is comparable to that generated by the HBAR in our device, thus it is important to extract the local static strain potential for each NV center of interest. From the previous discussion, we know that $V_{A_{1}}$ only leads to an overall shift of $E$ state energy. It can then be treated as part of the $\ket{^{3}A_{2}}\rightarrow \ket{^{3}E}$ transition energy. $V_{E_{1}}$ and $V_{E_{2}}$ can be calculated from measuring the $E$ state splitting and the polarization of the $\ket{E_{x}}$ and $\ket{E_{y}}$ dipole transition in the following way: The diamond lattice $E$ type strain Hamiltonian is shown in Eq.~(\ref{eq:strain}). Through diagonalization of the Jahn-Teller matrix, we find the eigenvalues $\Delta_{x}=\sqrt{V^{2}_{E_{1}}+V^{2}_{E_{2}}}$ and $\Delta_{y}=-\sqrt{V^{2}_{E_{1}}+V^{2}_{E_{2}}}$~\cite{JayichPRA}, which correspond to the $E$ state energy shift with respect to $\hbar\omega_{0}+V_{A_{1}}$. The new eigenstates are superpositions of $\{\ket{E_{x}}$ and $\ket{E_{y}}\}$, with an mixing angle given by  tan$2\theta=V_{E_{2}}/V_{E_{1}}$. The associated orthogonal dipole moment of the $\ket{E_{x}},\ket{E_{y}}$ states will effectively experience a clockwise rotation of $\theta$.

\begin{equation}
\label{eq:strain}
H_{\text{int}}-V_{A_{1}}=
\begin{pmatrix}
V_{E_{1}} & V_{E_{2}} \\ 
V_{E_{2}}& -V_{E_{1}}
\end{pmatrix}
=\begin{pmatrix}
\text{cos}\theta & -\text{sin}\theta \\ 
\text{sin}\theta & \text{cos}\theta
\end{pmatrix}
\begin{pmatrix}
\sqrt{V^{2}_{E_{1}}+V^{2}_{E_{2}} }& 0 \\ 
0& -\sqrt{V^{2}_{E_{1}}+V^{2}_{E_{2}}}
\end{pmatrix}
\begin{pmatrix}
\text{cos}\theta & \text{sin}\theta \\ 
-\text{sin}\theta & \text{cos}\theta
\end{pmatrix}
\end{equation}

Considering a [111] orientated NV center, with a strain perturbation along $Z$ ([001]), the symmetry of the NV center reduces from $C_{3v}$ down to $C_{1h}$. Symmetry breaking aligns the $x$ dipole and $y$ dipole respectively along [$\bar{1}\bar{1}2$] and [$1\bar{1}0$]. As shown in Eq.~(\ref{eq:strain}), extra perturbation from an $E_{2}$ interaction will rotate the dipole by $\theta$ (Fig.~\ref{fig:crystal}).

\begin{figure}[!htbp]
 \centering
\includegraphics{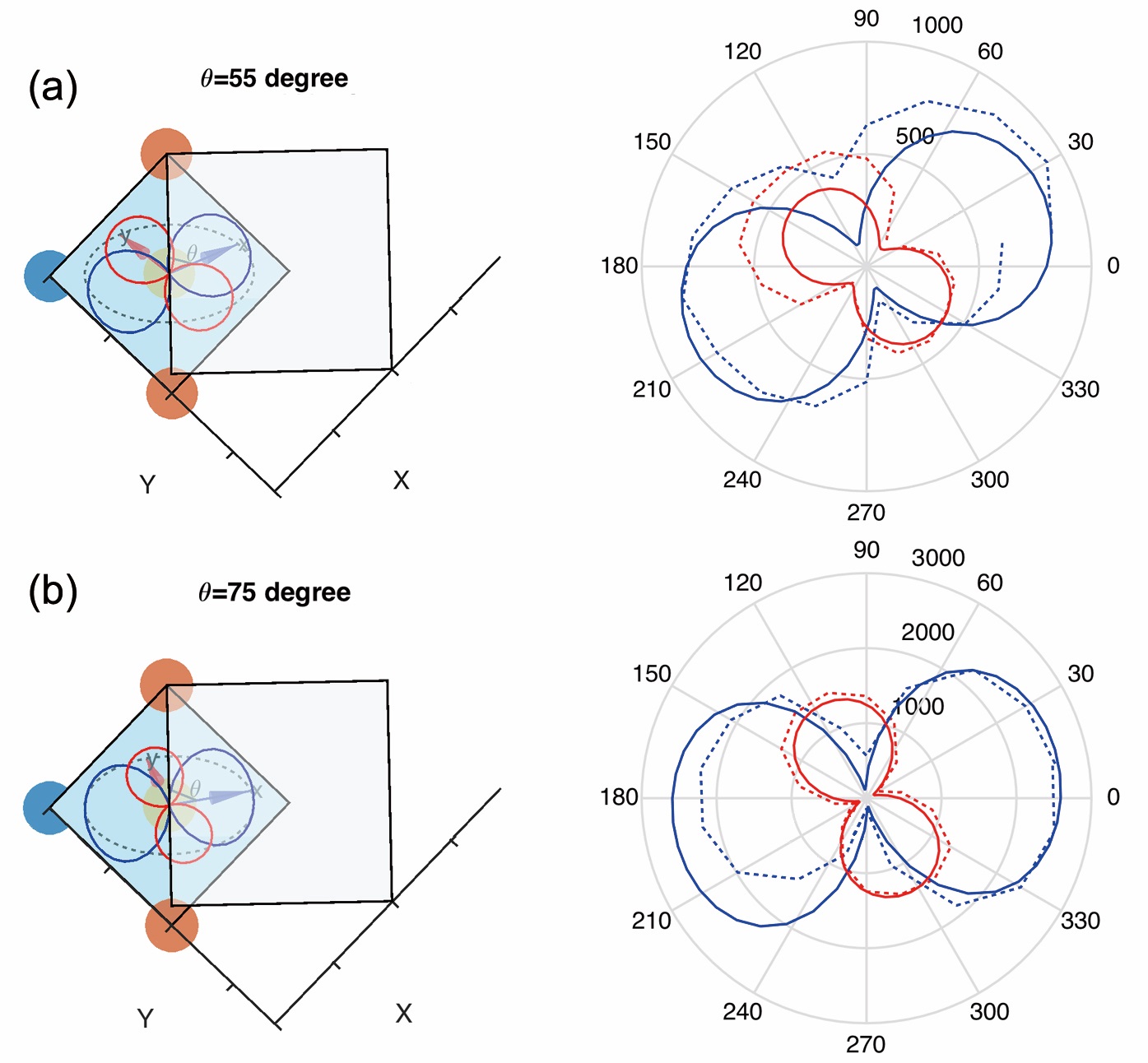}
\caption{Polarization data for two different NV centers, plotted according to Fig.~\ref{fig:crystal}. Dotted lines are experimental data and solid lines are a corresponding fit. (a) NV1 (b) NV2.}
\label{fig:polar}
\end{figure}

Experimental polarization measurements are shown in Fig.~\ref{fig:polar}. The mixing angles $\theta$ are extracted from the fitting after taking into account the laser saturation~\cite{JayichPRA}.

\section{Simulation and analysis}\label{Ans}

\subsection{Quantum master equation simulation}

Experimentally, with a weak probe laser of frequency $\omega_{l}$, one can spectroscopically resolve the $\ket{^{3}A_{2}}\rightarrow \ket{^{3}E}$ optical transition, in which the zero-spin transition frequency is determined by $H_{0}\ket{E}=\hbar\omega_{0}\ket{E}$. As seen in Fig.~\ref{fig:level}(b), the ground state electron energy is stable against strain perturbations, due to the anti-symmetrized orbital configuration. Strain induced splitting and orbital dynamics in the NV center excited states can then be easily detect from the photo-luminescence  spectrum.

For simplicity, we write state $\ket{^{3}A_{2}, m_{s}=0}$, $\ket{E_{x}}$ and $\ket{E_{y}}$ as $\ket{g}$, $\ket{x}$ and $\ket{y}$, $A_{1}\sigma_{0}\rightarrow\mathcal{A}$ and $2B\sigma_{0}\rightarrow\mathcal{E}_{1}$. Under the rotating wave approximation for the dipole interaction,
\begin{equation}
\label{eq:master}
\begin{split}
H=&(\hbar\omega_{0}+V_{A1}+\mathcal{A}\text{cos}\omega_{m}t)(\ket{x}\bra{x}+\ket{y}\bra{y})
\\&+(V_{E_{1}}+\mathcal{E}_{1}\text{cos}\omega_{m}t)(\ket{x}\bra{x}-\ket{y}\bra{y})+V_{E_{2}}(\ket{x}\bra{y}+\ket{y}\bra{x})
\\&+\frac{\hbar}{2}\Omega(\ket{g}\bra{x} e^{i\omega_{l}t}+\ket{x}\bra{g}e^{-i\omega_{l}t})
+\frac{\hbar}{2}\Omega(\ket{g}\bra{y} e^{i\omega_{l}t}+\ket{y}\bra{g}e^{-i\omega_{l}t}).
\end{split}
\end{equation}
Transforming into the laser rotating frame, we have
\begin{equation}
\label{eq:master0}
\begin{split}
H=&(-\hbar\Delta+\mathcal{A}\text{cos}\omega_{m}t)(\ket{x}\bra{x}+\ket{y}\bra{y})
\\&+(V_{E_{1}}+\mathcal{E}_{1}\text{cos}\omega_{m}t)(\ket{x}\bra{x}-\ket{y}\bra{y})+V_{E_{2}}(\ket{x}\bra{y}+\ket{y}\bra{x})
\\&+\frac{\hbar}{2}\Omega(\ket{g}\bra{x} +\ket{x}\bra{g})
+\frac{\hbar}{2}\Omega(\ket{g}\bra{y} +\ket{y}\bra{g}),
\end{split}
\end{equation}
where $\Delta=\omega_{l}-\omega_{0}-V_{A1}/\hbar$ is the laser detuning.

For the simulation, we dynamically evolve of the system from the initial ground state $\ket{g}$, using quantum master equation of Lindblad form:

\begin{equation}
\label{eq:master1}
\begin{split}
\dot{\rho}&=-i[H,\rho]+\mathcal{L}_{\Gamma_{x}}\rho+\mathcal{L}_{\Gamma_{y}}\rho,
\\
\rho_{t=0}&=\ket{g}\bra{g},
\end{split}
\end{equation}
where the dissipator $\mathcal{L}$ describes the optical decay of the excited states:

\begin{equation}
\label{eq:master2}
\begin{split}
\mathcal{L}_{\Gamma_{x}}\rho&=\Gamma(\ket{g}\bra{x}\rho \ket{x}\bra{g}-\frac{1}{2}\ket{x}\bra{x}\rho-\frac{1}{2}\rho \ket{x}\bra{x}),
\\
\mathcal{L}_{\Gamma_{y}}\rho&=\Gamma(\ket{g}\bra{y}\rho \ket{y}\bra{g}-\frac{1}{2}\ket{y}\bra{y}\rho-\frac{1}{2}\rho \ket{y}\bra{y}).
\end{split}
\end{equation}
$1/\Gamma$ is the excited state lifetime of NV center.

\subsection{Phonon sidebands from photon-phonon coupling }\label{ppcoupling}

\begin{figure}[!htbp]
\centering
\includegraphics{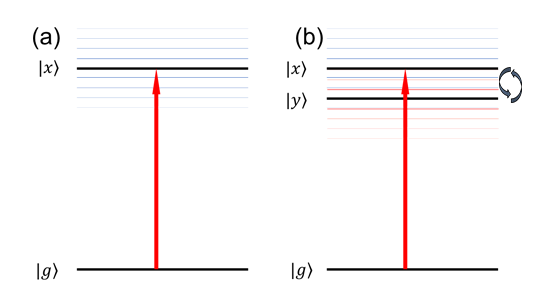}
\caption{(a) Phonon sidebands generated upon $\ket{x}$ state modulation. (b) Phonon induced orbital transitions due to $E$ type electron-phonon coupling.}
\label{fig:sidebands}
\end{figure}

For sufficiently large static strain splitting ($2\Delta_{x}\gg \hbar\omega_{m},\mathcal{A},\mathcal{E}_{1}$), $\ket{E_{x}}$ and $\ket{E_{y}}$ are decoupled, the three level Hamiltonian described by Eq.~(\ref{eq:master0}) distills to a two level system formed by $\{\ket{x},\ket{g}\}$ (or $ \{\ket{y},\ket{g}\}$) with wave function $\psi=c_{g}\ket{g}+c_{x}\ket{x}$. Having the AC strain perturbation modulate the $E$ state, the resulting Schrodinger's equation is

\begin{equation}
\label{eq:schrodinger}
i\hbar\frac{\mathrm{d} }{\mathrm{d} t}\begin{pmatrix}
c_{g}\\ 
c_{x}
\end{pmatrix}
=\begin{pmatrix}
0 & \hbar\Omega/2 \\ 
 \hbar\Omega/2& -\hbar\Delta+\mathbf{A}\text{cos}\omega_{m} t
\end{pmatrix}
\begin{pmatrix}
c_{g}\\ 
c_{x}
\end{pmatrix},
\end{equation}
where $\mathbf{A}=\mathcal{A}\pm\mathcal{E}_{1}$. The solution can be solved through Floquet theory as
\begin{equation}
\label{eq:schrodinger2}
\begin{split}
c_{g}&=e^{-iqt}\sum_{n}^{ }c_{g,n}e^{-in\omega_{m}t},
\\
c_{x}&=e^{-iqt}\sum_{n}^{ }c_{x,n}e^{-in\omega_{m}t},
\end{split}
\end{equation}
where $q$ is the quasi-energy, $c_{g,n}$ and $c_{x,n}$ are probability amplitude of the Floquet states $\ket{g,m}$ and $\ket{x,n}$, where $m$ and $n$ are coherent phonon numbers of the dressed states. Diagonalization of the Floquet Hamiltonian,
\begin{equation}
\label{eq:FloquetHam}
H_F=\hbar
\begin{pmatrix}
\ddots  & \vdots  & \vdots &\vdots  &\vdots  &\vdots  &\vdots  &\vdots  & \\ 
\hdots&  \omega_{m}&0  & 0 &\hdots  &\frac{\Omega}{2}J_{0}  &\frac{\Omega}{2}J_{1}  &\frac{\Omega}{2}J_{2}  & \hdots \\ 
\hdots &0  &0  & 0 &\hdots   & \frac{\Omega}{2}J_{-1} & \frac{\Omega}{2}J_{0} &\frac{\Omega}{2}J_{1}  & \hdots \\ 
 \hdots& 0&0  & -\omega_{m} &\hdots   &\frac{\Omega}{2}J_{-2}  & \frac{\Omega}{2}J_{-1} &\frac{\Omega}{2}J_{0}  & \hdots \\ 
 \hdots&\vdots  &\vdots  &\vdots  & \ddots & \vdots & \vdots &\vdots  & \hdots \\ 
 \hdots& \frac{\Omega}{2}J_{0} &  \frac{\Omega}{2}J_{1}&\frac{\Omega}{2}J_{2}  & \hdots &-\Delta+\omega_{m}  &0  &0  & \hdots \\ 
 \hdots&  \frac{\Omega}{2}J_{-1}& \frac{\Omega}{2}J_{0} &\frac{\Omega}{2}J_{1}  &\hdots  &0  &-\Delta  &0  & \hdots \\ 
 \hdots& \frac{\Omega}{2}J_{-2} &\frac{\Omega}{2}J_{-1}  &\frac{\Omega}{2}J_{0}  & \hdots &0  &0  &-\Delta-\omega_{m}  & \hdots \\ 
 &  \vdots&\vdots  &\vdots  &\vdots  &\vdots  &\vdots  &\vdots  & 
\end{pmatrix}
\begin{matrix*}[l]
\vdots \\ 
\leftarrow \ket{g,m=1}\\ 
\leftarrow \ket{g,m=0}\\ 
\leftarrow\ket{g,m=-1}\\ 
\vdots\\ 
\leftarrow \ket{x,n=1}\\ 
\leftarrow \ket{x,n=0}\\ 
\leftarrow \ket{x,n=-1}\\ 
\vdots
\end{matrix*}
\end{equation}
yields quasienergy levels separated evenly in frequency by $\omega_{m}$, which corresponds to the generation of phonon sidebands in the resolved sideband regime, $\omega_{m}>\Gamma$. The off-diagonal terms $\frac{\Omega}{2}J_{n-m}(\mathbf{A}/\hbar\omega_{m})\ket{g,m}\bra{x,n}$, characterize the Bessel function form of photon-phonon coupling between these states~\cite{theorypaperSQC}.

\begin{figure}[!ht]
\centering
\includegraphics{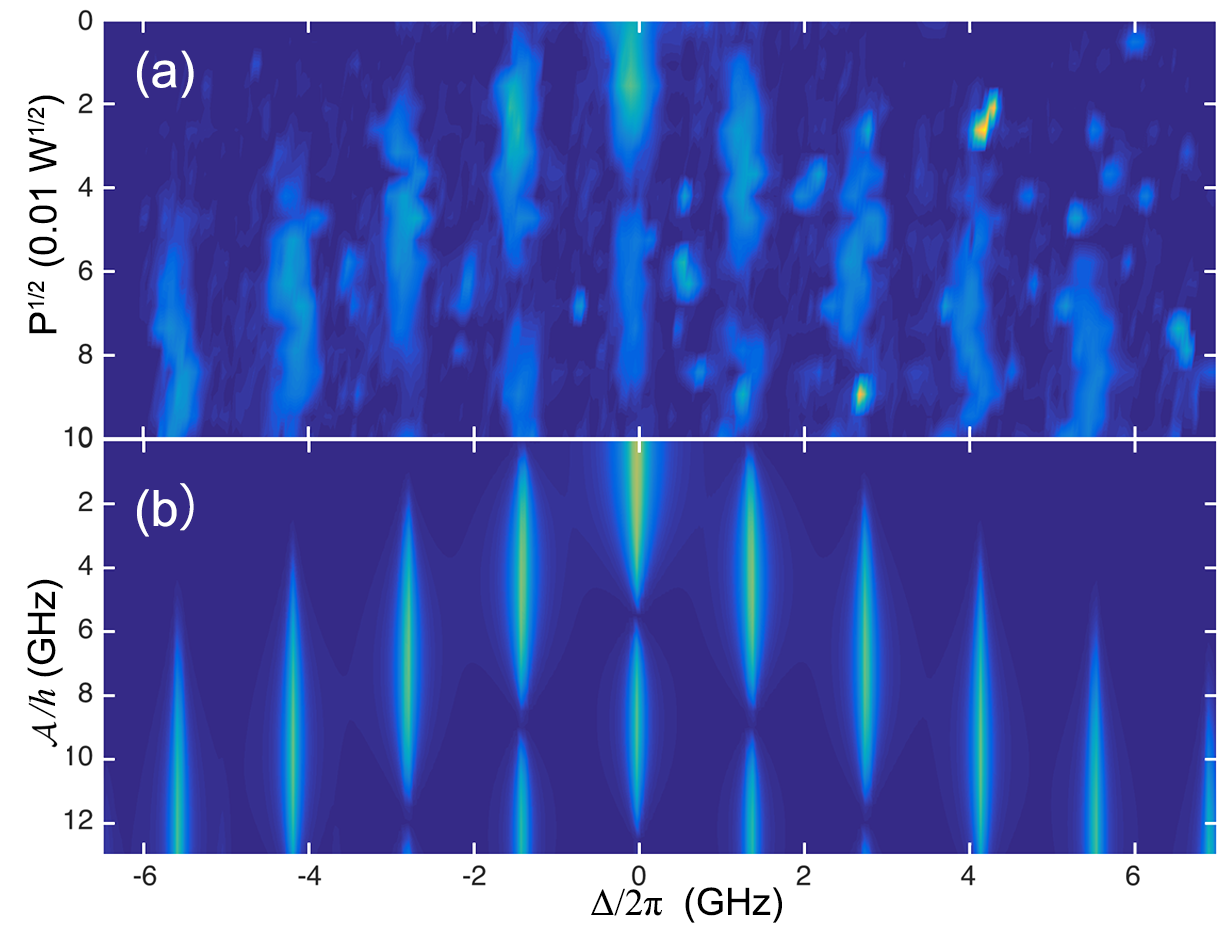}
\caption{PLE spectrum of a large $E$ state splitting ($2\Delta_{x}/\hbar$=23 GHz) NV center (NV4) with increasing mechanical driving amplitude. Laser detuning $\Delta$ is set around $\ket{^{3}A_{2}}\rightarrow \ket{E_{x}}$ transition. (a) Mechanical driving at 1.3844 GHz. (b) Quantum master equation simulation result from Eq.(\ref{eq:master0}), simulation parameters: $\Omega/2\pi=0.2$ GHz, $\Gamma/2\pi$= 60 MHz, $\omega_{m}/2\pi$=1.3844 GHz.}
\label{fig:sidebandsexp}
\end{figure}

In this regime [Fig.\ref{fig:sidebands}(a)], both $A$ and $E_{1}$ types of modulation contribute to the presence of phonon-sidebands and it is hard to distinguish the two. The relevant experimental data is present in Fig.~\ref{fig:sidebandsexp} for an NV center (NV4) with 23 GHz static strain $\ket{E}$ state splitting.

\subsection{Phonon-driven orbital transition from orbital-phonon coupling }

As the strain modulation frequency approaches the local strain splitting ($2\Delta_{x}\sim\hbar\omega_{m}$), $E_{1}$ modulation $\mathcal{E}_{1}cos\omega_{m}t$ along with $V_{E_{2}}$ can lead to phonon-driven orbital transitions between $\ket{E_{x}},\ket{E_{y}}$ [Fig.~\ref{fig:sidebands}(b)]. For simplicity, we limit our discussion to the excited-state manifold $\psi=c_{x}\ket{x}+c_{y}\ket{y}$. The Hamiltonian then reduces from Eq.~(\ref{eq:master0}) to 

\begin{equation}
\label{eq:RWT}
\begin{split}
H&=
\begin{pmatrix}
V_{E_{1}}+(\mathcal{A}+\mathcal{E}_{1})\text{cos}\omega_{m}t & V_{E_{2}}\\ 
V_{E_{2}} & -V_{E_{1}}+(\mathcal{A}-\mathcal{E}_{1})\text{cos}\omega_{m}t 
\end{pmatrix}.
\end{split}
\end{equation}
We will treat the strain modulation as a perturbation to the Hamiltonian of the intrinsic local static strain in the rotated basis as discussed in Eq.~(\ref{eq:strain}):
\begin{equation}
\label{eq:RotatedBasis}
\begin{split}
\begin{pmatrix}
\psi^{R}_{x}\\ 
\psi^{R}_{y}
\end{pmatrix}
=
U_{R}
\begin{pmatrix}
\psi_{x}\\ 
\psi_{y}
\end{pmatrix},
\quad \textrm{and} \quad 
U_{R}=\begin{pmatrix}
\text{cos}\theta &\text{sin}\theta \\ 
 -\text{sin}\theta& \text{cos}\theta
\end{pmatrix},
\end{split}
\end{equation}
where $\theta$ is the mixing angle, defined by tan$2\theta=V_{E_{2}}/V_{E_{1}}$ from previous discussion in section~\ref{intrinsic}. In the rotated basis, Eq.(\ref{eq:RWT}) becomes
\begin{equation}
\label{eq:RotatedBasisH}
H_{R}=U_{R}HU^{-1}_{R}
=\begin{pmatrix}
\Delta_{x}+(\mathcal{A}+\mathcal{E}_{1}\text{cos}2\theta)\text{cos}\omega_{m}t 
& -\mathcal{E}_{1}\text{sin}2\theta\text{cos}\omega_{m}t\\ 
 -\mathcal{E}_{1}\text{sin}2\theta\text{cos}\omega_{m}t
& -\Delta_{x}+(\mathcal{A}-\mathcal{E}_{1}\text{cos}2\theta)\text{cos}\omega_{m}t
\end{pmatrix}.
\end{equation}
The physical effect of the off-diagonal term will become more evident as we go into the rotating frame of the modulation frequency $\omega_{m}$ by applying the transformation:
\begin{equation}
\label{eq:RWAH1}
\begin{pmatrix}
\tilde{\psi}^{R}_{x}\\ 
\tilde{\psi}^{R}_{y}
\end{pmatrix}
=
U^{\dagger}
\begin{pmatrix}
\psi^{R}_{x}\\ 
\psi^{R}_{y}
\end{pmatrix},
\quad \textrm{and} \quad 
U^{\dagger}=\begin{pmatrix}
e^{i\omega_{m}t} &0  \\ 
 0& 1
\end{pmatrix}
\end{equation}

\begin{equation}
\label{eq:RWAH2}
\begin{split}
\tilde{H}_{R}&=U^{\dagger}H_{R}U-iU^{\dagger}\frac{\mathrm{d} U}{\mathrm{d} t}
\\
&=\begin{pmatrix}
-\hbar\omega_{m}+\Delta_{x}+(\mathcal{A}+\mathcal{E}_{1}\text{cos}2\theta)\text{cos}\omega_{m}t 
& -\mathcal{E}_{1}\text{sin}2\theta\\ 
-\mathcal{E}_{1}\text{sin}2\theta
& -\Delta_{x}+(\mathcal{A}-\mathcal{E}_{1}\text{cos}2\theta)\text{cos}\omega_{m}t
\end{pmatrix},
\end{split}
\end{equation}
where in the last step, we have used the rotating wave approximation and neglected the counter-rotating terms. 

It's now evident that the off-diagonal terms in Eq.~(\ref{eq:RWAH2}) couple the two $\ket{E}$ orbital states $\{\psi^{R}_{x},\psi^{R}_{y}\}$ and induce phonon-driven orbital transitions when the resonance condition, $\hbar\omega_{m}=2\Delta_{x}$, is met. There is an asymmetry in the diagonal modulation, which will be detectable in experiment as asymmetric phonon sidebands if $ \mathcal{E}_{1}\text{cos}2\theta $ is comparable to $\mathcal{A}$.

However, Eq.~(\ref{eq:RWAH2}) doesn't reveal the multi-phonon orbital transitions seen in our experiments through $E_{1}$ modulation, $ \mathcal{E}_{1}\text{cos}2\theta $ term in Eq.~(\ref{eq:RWAH2}). To do so, we can apply an unitary polaron transformation~\cite{JCP,PRA2007} $\ket{\tilde{\psi_{R}}}=U_{p}\ket{\psi_{D}}$, with $U_{p}=\text{exp}[i(\frac{\mathcal{E}_{1}}{\hbar\omega_{m}}\text{cos}2\theta\text{sin}\omega_{m}t)\hat{\sigma}_{z}]$. The Hamiltonian in the displaced oscillator basis $\ket{\psi_{D}}$~\cite{2010PRB} then becomes,

\begin{equation}
\label{eq:RWT2SI}
\begin{split}
{H}'&=i\frac{ \mathrm{d} U_{p}^{\dagger} }{\mathrm{d} t}U_{p}+U_{p}^{\dagger}\tilde{H}_{R}U_{p}
\\
&=\begin{pmatrix}
-\hbar\omega_{m}+\Delta_{x}+\mathcal{A}\text{cos}\omega_{m}t & -\mathcal{E}_{1}\text{sin}2\theta e^{2i\frac{\mathcal{E}_{1}}{\hbar\omega_{m}}\text{cos}2\theta\text{sin}\omega_{m}t}  \\ 
-\mathcal{E}_{1}\text{sin}2\theta e^{-2i\frac{\mathcal{E}_{1}}{\hbar\omega_{m}}\text{cos}2\theta\text{sin}\omega_{m}t} & -\Delta_{x}+\mathcal{A}\text{cos}\omega_{m}t
\end{pmatrix}
\\
&=
\begin{pmatrix}
-\hbar\omega_{m}+\Delta_{x}+\mathcal{A}\text{cos}\omega_{m}t  & -\mathcal{E}_{1}\text{sin}2\theta\sum\limits_{n} J_{n}(2\frac{\mathcal{E}_{1}}{\hbar\omega_{m}}\text{cos}2\theta)e^{-in\omega_{m}t}\\ 
-\mathcal{E}_{1}\text{sin}2\theta\sum\limits_{n} J_{n}(2\frac{\mathcal{E}_{1}}{\hbar\omega_{m}}\text{cos}2\theta)e^{in\omega_{m}t} & -\Delta_{x}+\mathcal{A}\text{cos}\omega_{m}t 
\end{pmatrix}.
\end{split}
\end{equation}
Note that the off diagonal term represents the orbital-phonon coupling and has a similar form to photon-phonon coupling in trapped ion systems~\cite{reviewModernPhysics}. As $\mathcal{E}_{1}\rightarrow0$, Eq.~(\ref{eq:RWT2SI}) reduces to Eq.~(\ref{eq:RWAH2}). When the resonance condition is satisfied such that $(n+1)\hbar\omega_{m}=2\Delta_{x}$, the dominant term in orbital-phonon coupling is $-\mathcal{E}_{1}\text{sin}2\theta J_{n}(2\frac{\mathcal{E}_{1}}{\hbar\omega_{m}}\text{cos}2\theta)$, which is effectively the phonon Rabi frequency, and $2\frac{\mathcal{E}_{1}}{\hbar\omega_{m}}\text{cos}2\theta$ is the effective Lamb-Dicke parameter. When $\hbar\omega_{m}>\mathcal{E}_{1}\text{sin}2\theta$~\cite{PRA2007}, approximating the Hamiltonian near or on an $(n+1)$-phonon resonance gives,

\begin{equation}
\label{eq:RWT3}
\begin{split}
{H}'_{(n+1)\hbar\omega_{m}\sim2\Delta_{x}}
&=\mathcal{A}\text{cos}\omega_{m}t+
\begin{pmatrix}
-\hbar\omega_{m}+\Delta_{x} & -\mathcal{E}_{1}\text{sin}2\theta J_{n}(2\frac{\mathcal{E}_{1}}{\hbar\omega_{m}}\text{cos}2\theta)e^{-in\omega_{m}t}\\ 
-\mathcal{E}_{1}\text{sin}2\theta J_{n}(2\frac{\mathcal{E}_{1}}{\hbar\omega_{m}}\text{cos}2\theta)e^{in\omega_{m}t} & -\Delta_{x}
\end{pmatrix}
\\
&\xrightarrow{RWA}
\mathcal{A}\text{cos}\omega_{m}t+
\begin{pmatrix}
0 & -\mathcal{E}_{1}\text{sin}2\theta J_{n}(2\frac{\mathcal{E}_{1}}{\hbar\omega_{m}}\text{cos}2\theta)\\ 
-\mathcal{E}_{1}\text{sin}2\theta J_{n}(2\frac{\mathcal{E}_{1}}{\hbar\omega_{m}}\text{cos}2\theta) & 0
\end{pmatrix},
\end{split}
\end{equation}
in the last step of which we have applied RWA.

\begin{figure}[!ht]
\centering
\includegraphics{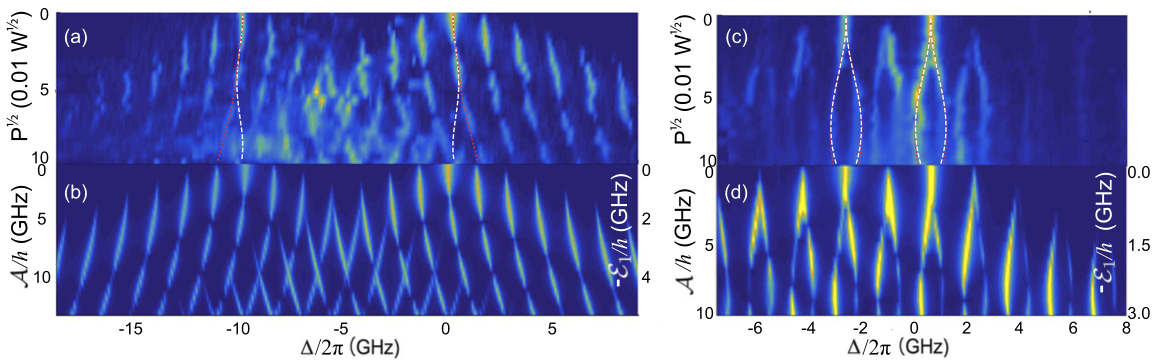}
\caption{(a) PLE spectrum of an NV center (NV5) with $E$ state splitting 9.8 GHz ($\simeq7 \omega_{m}$), driven at $\omega_{m}=$1.3844 GHz. (c) PLE spectrum of a $2\Delta_{x}=3.2$ GHz ($\simeq2 \omega_{m}$) splitting NV (NV2) driven at $\omega_{m}=1.6$ GHz. The simulation results of the two NVs are shown in (b) and (d) respectively. The two NV centers show phonon Rabi splitting, described approximately by $\sqrt{\sum_{n+1=1}^{8}S(n+1)}$ [Eq.~(\ref{eq:RS})] and indicated by the red dotted lines. For NV5, single phonon process contributes the most to the Rabi splitting in the driving amplitude range $\mathcal{E}_{1}/h<5$ GHz [white dashed line in (a)], while the higher order phonon process participate more for the range $\mathcal{E}_{1}/h>5$ GHz. For NV2, the resonant 2-phonon process dominates the dressed state behavior, and contributes the most to the total Rabi splitting, which can be described by $\mathcal{E}_{1}\text{sin}2\theta J_{1}(2\frac{\mathcal{E}_{1}}{\hbar\omega_{m}}\text{cos}2\theta)$, shown as the white dashed lines in (c).}
\label{fig:phonontransition}
\end{figure}

Eq.~(\ref{eq:RWT3}) substantiates the existence of $(n+1)$ multi-phonon driven orbital transition under $E_{1}$ modulation [Fig.~\ref{fig:sidebands}]. The orbital-phonon coupling as well as the phonon Rabi splitting follow a Bessel function form as the modulation strength increases, which corresponds to the energy shift of transition lines in our experiment[Fig.~\ref{fig:phonontransition}]. The first term in Eq.~(\ref{eq:RWT2SI}) will generate sidebands in the same way as discussed in section~\ref{ppcoupling} due to photon-phonon coupling.

\begin{figure}[!ht]
\centering
\includegraphics{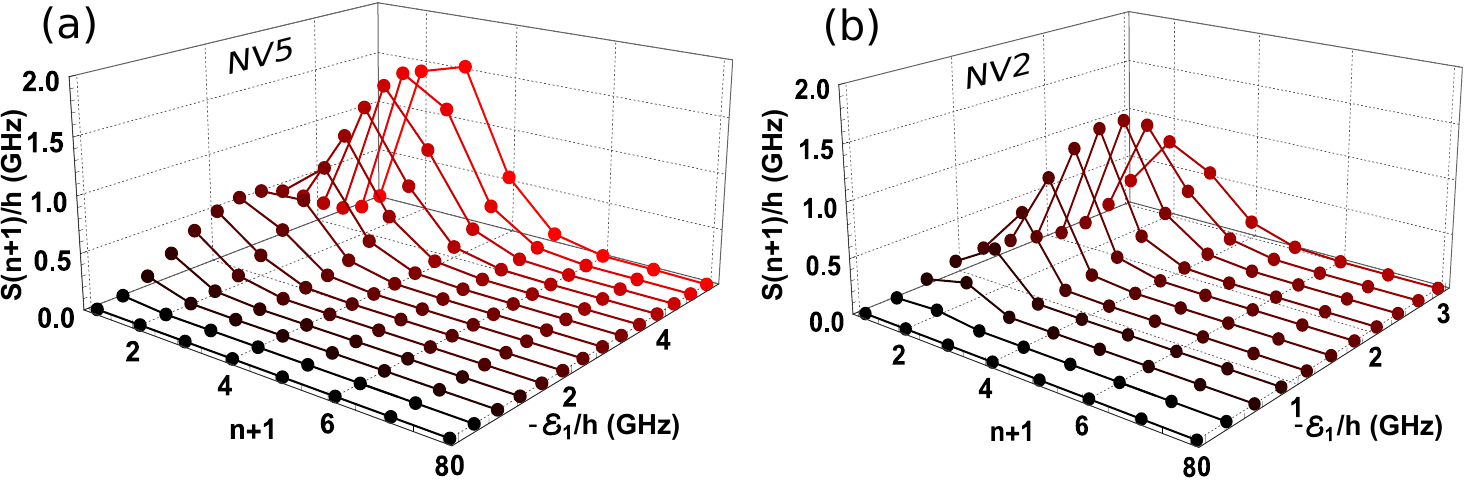}
\caption{Rabi splitting contribution from each order phonon interaction $S(n+1)$ plotted as a function of $\mathcal{E}_{1}$ amplitude for (a) NV5 and (b) NV2. The lower order phonon interactions, $n+1<4$, contribute the most to NV5 dressed state, while $n+1=2$ phonon interaction dominates the dressed state behavior of NV2.}
\label{fig:PhononContribution}
\end{figure}

To quantify the phonon-induced Rabi splitting, we take measurements on NV5 and NV2, where the static strain splitting matches 7 and 2 times the phonon frequency, respectively (Fig.~\ref{fig:phonontransition}). Due to simultaneous multi-phonon process, it is hard to diagonalize Eq.~(\ref{eq:RWT2SI}). Instead, we plot the Rabi splitting contribution $S(n+1)$ from each order of phonon driving, 
\begin{equation}
\label{eq:RS}
S(n+1)=\sqrt{[2 \Delta_{x} - (n + 1)\hbar\omega_{m}]^2 +[2\mathcal{E}_{1}\text{sin}2\theta J_{n}(2\frac{\mathcal{E}_{1}}{\hbar\omega_{m}}\text{cos}2\theta)]^{2}} -[2\Delta_{x}-(n+1)\hbar\omega_{m}],
\end{equation}
as shown in Fig.~\ref{fig:PhononContribution}. The distribution of Rabi contribution from each order of phonon interaction varies as a function of $\mathcal{E}_{1}$, while the low-order phonon modes contribute the most. Due to the large static strain splitting of NV5, high order phonon interactions, including the $n+1$=7 resonant phonon mode, are suppressed, and $n+1=1\sim3$ phonon processes dominant the dressed state behavior within our mechanical driving range. In contrast to NV5, 2-phonon processes contribute the most to the Rabi splitting and dominate the dressed state for NV2 throughout the measurement range.

We plot and overlay the Rabi splitting from $n+1=1$ and $n+1=2$ phonon interactions respectively on the dressed state PLE spectra for NV5 and NV2 [white dashed lines in Fig.~\ref{fig:phonontransition}(a, c)]. To include the Rabi contribution from other orders of phonon interactions, we also plot the root sum square of Rabi splittings for up to $n+1=8$ orders of phonon interaction, $\sqrt{\sum_{n+1=1}^{8}S(n+1)}$ [Red dotted lines in Fig.~\ref{fig:phonontransition}(a, c)]. The agreement of experimental data and our analysis again supports the existence of (n+1) multi-phonon driven orbital transition under $E_{1}$ modulation.

\section{$E_{2}$ interaction from stray stress waves}

When the applied stress is not perfectly aligned along the diamond crystal axis, one sees from Eq.~(\ref{eq:JT4}) that a finite $E_{2}$ interaction can be generated [Eq.(\ref{eq:E2})].

\begin{equation}
\label{eq:E2}
\begin{split}
H&=
\begin{pmatrix}
V_{E_{1}}+(\mathcal{A}+\mathcal{E}_{1})\text{cos}\omega_{m}t & V_{E_{2}}+\mathcal{E}_{2}\text{cos}\omega_{m}t\\ 
V_{E_{2}}+\mathcal{E}_{2}\text{cos}\omega_{m}t & -V_{E_{1}}+(\mathcal{A}-\mathcal{E}_{1})\text{cos}\omega_{m}t 
\end{pmatrix}
\end{split}
\end{equation}

Note that in the rotated dipole basis, $E_{2}$ modulation can result in a similar effect as $E_{1}$ modulation, and this could lead to a suppression of the $E_{1}$ interaction by $\mathcal{E}_{2}\text{cos}2\theta$, as shown in the diagonal term in Eq.~(\ref{eq:E22}). As a result, the ratio of $A_{1}$ and $E_{1}$ interactions measured in the experiment could deviate from the expected value: $\mathcal{E}_{1}/\mathcal{A}<2B/A_{1}$ 

\begin{equation}
\label{eq:E22}
H_{R}=U_{R}HU^{-1}_{R}
=\begin{pmatrix}
\Delta_{x}+(\mathcal{A}+\mathcal{E}_{1}\text{cos}2\theta+\mathcal{E}_{2}\text{sin}2\theta)\text{cos}\omega_{m}t 
& (-\mathcal{E}_{1}\text{sin}2\theta+\mathcal{E}_{2}\text{cos}2\theta)\text{cos}\omega_{m}t\\ 
 (-\mathcal{E}_{1}\text{sin}2\theta+\mathcal{E}_{2}\text{cos}2\theta)\text{cos}\omega_{m}t
& -\Delta_{x}+(\mathcal{A}-\mathcal{E}_{1}\text{cos}2\theta-\mathcal{E}_{2}\text{sin}2\theta)\text{cos}\omega_{m}t
\end{pmatrix}
\end{equation}

\section{8-level Hamiltonian and simulation}
Simulating the spin-1 transitions requires the full description of the 6-level excited state coupling to the ground states. After RWA, the Hamiltonian can be written as Eq.~(\ref{eq:8level}), in the basis of $\{\ket{A_{1}}, \ket{A_{2}}, \ket{E_{x}}, \ket{E_{y}}, \ket{E_{1}}, \ket{E_{2}}, \ket{^3A_{2}, \abs{m_{s}}=1}, \ket{^3A_{2}, m_{s}=0}\}$.

\begin{equation}
\small
\label{eq:8level}
\begin{split}
H=
&\begin{pmatrix}
\lambda_{z}+\frac{1}{3}D-{\Delta}' & 0 & 0 & 0 & V_{E_{1}} &  V_{E_{2}} &\frac{\hbar\Omega}{2}&0\\ 
 0&\lambda_{z}+\frac{1}{3}D+{\Delta}'  &  0&0  &  V_{E_{2}}&  -V_{E_{1}}&\frac{\hbar\Omega}{2} &0\\ 
 0& 0 & -\frac{2}{3}D+V_{E_{1}} & V_{E_{2}} & 0 & i\lambda_{xy} & 0&\frac{\hbar\Omega}{2}\\ 
 0& 0 &  V_{E_{2}}&-\frac{2}{3}D+V_{E_{1}}   & \lambda_{xy} &0  &0 &\frac{\hbar\Omega}{2}\\ 
V_{E_{1}} & V_{E_{2}} & 0 &  \lambda_{xy}&  -\lambda_{z}+\frac{1}{3}D&  0&\frac{\hbar\Omega}{2} &0\\ 
 V_{E_{2}}&  -V_{E_{1}}& -i\lambda_{xy} &0  & 0 &-\lambda_{z}+\frac{1}{3}D  &\frac{\hbar\Omega}{2} &0\\ 
 \frac{\hbar\Omega}{2}&\frac{\hbar\Omega}{2}  & 0 &0  & \frac{\hbar\Omega}{2} &  \frac{\hbar\Omega}{2}&\Delta+D_{0}+V_{\parallel } &\frac{\hbar\Omega_{\text{mw}}}{2}\\
0 &  0& \frac{\hbar\Omega}{2} & \frac{\hbar\Omega}{2} &0  &0  & \frac{\hbar\Omega_{\text{mw}}}{2}&\Delta
\end{pmatrix}
\\&+
\begin{pmatrix}
 \mathcal{A} &0  & 0 &  0& \mathcal{E}_{1}&  0&  0&0 \\ 
 0& \mathcal{A}&  0 & 0 & 0 &  -\mathcal{E}_{1}&0  &0 \\ 
 0& 0  &\mathcal{A}+\mathcal{E}_{1}& 0  & 0 &0  &  0&0 \\ 
 0& 0 & 0  & \mathcal{A}-\mathcal{E}_{1}&0  &0  &  0&0 \\ 
 \mathcal{E}_{1}&  0&  0 &0  &\mathcal{A} &0  & 0 &0 \\ 
 0& -\mathcal{E}_{1} &0  & 0  & 0 &\mathcal{A}& 0 & 0\\ 
 0&  0&  0&  0& 0 & 0 &  0& 0\\ 
 0& 0 &0  &0  &0  & 0 & 0 &0 
\end{pmatrix}\text{cos}\omega_{m}t ,
\end{split}
\end{equation}
where $\lambda_{z}/h=5.3$ GHz is the excited state spin-orbit interaction, $D/h$=1.42 GHz is the spin-spin interaction, ${\Delta}'/h$= 1.55 GHz, $\lambda_{xy}/h$=0.2 GHz~\cite{A.Batalov}, 
$D_{0}/h\simeq$ 2.877 GHz is the ground state spin-spin interaction splitting at 10K, $V_{\parallel}$ is the strain induced ground state spin energy shifting, and $\Omega_{\text{mw}}$ is the ground state spin magnetic driving Rabi frequency.

\begin{figure}[!ht]
\centering
\includegraphics{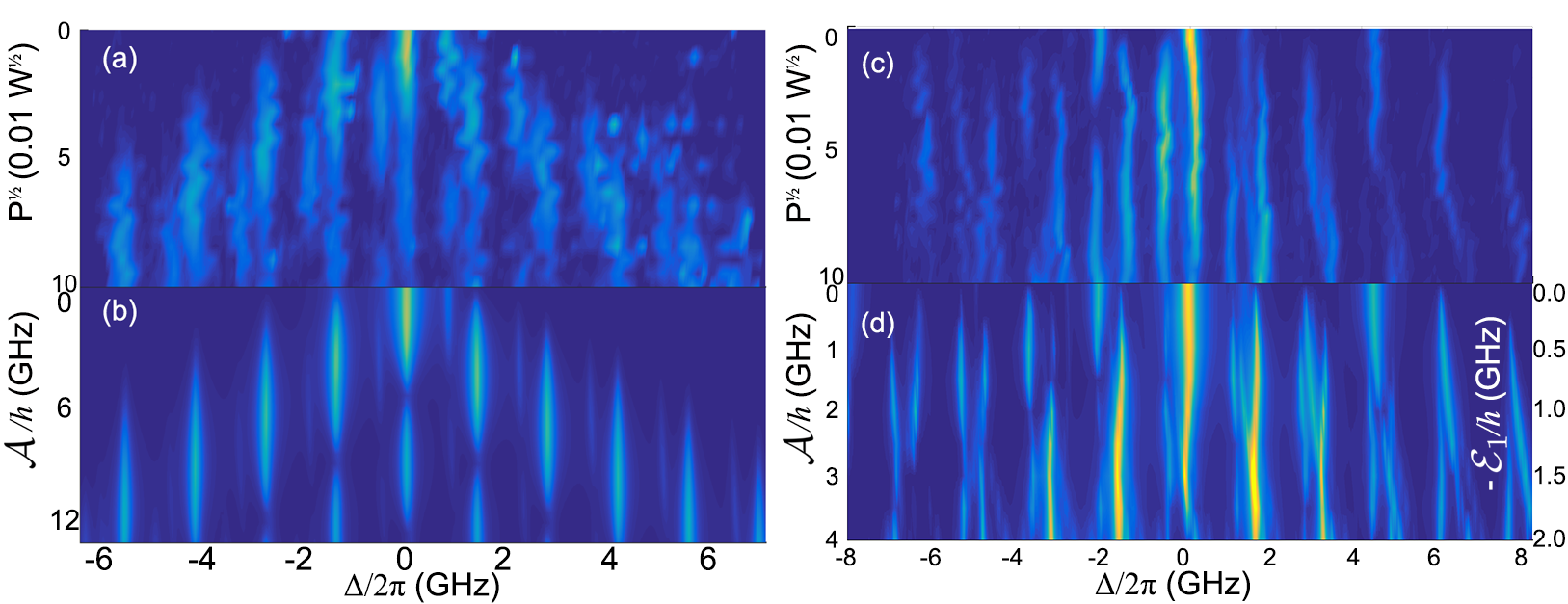}
\caption{(a) and (b) show the PLE dressed state spectra of NV4 and NV3, with 23 GHz and 2.1 GHz $E$ state static strain splitting, respectively. Magnetic microwave driving ground state spin at 2.876 GHz enables optical visibility of spin-1 transition. The detuning is zeroed at $\ket{^{3}A_{2}, m_{s}=0}$ to $\ket{E_{x}}$ transition. (c) and (d) are the simulation result based on the 8-level Hamiltonian in Eq.~(\ref{eq:8level}). }
\label{fig:8level}
\end{figure}

We dynamically evolve the density matrix from a mixed ground state (prepared via CW magnetic microwave driving at 2.877 GHz), $\rho=\frac{1}{2}\ket{g,1}\bra{g,1}+\frac{1}{2}\ket{g,0}\bra{g,0}$, through a quantum master equation similar as discussed above in Eq.~(\ref{eq:master1}-\ref{eq:master2}), with the simplified assumption that the optical decay of all excited states have the same rate $\Gamma$ (Decay through the meta-stable singlet states through inter-system crossing process has been neglected). As seen from Fig.~\ref{fig:8level}, the simulation agrees well with the experimental data. The 8-level Hamiltonian simulation also agrees well with the result from Eq.~(\ref{eq:master}-\ref{eq:master2}).

\section{Applications}

\subsection{Resonator Characterization through sideband transition}

As we see in section~\ref{Ans}, the sideband transitions from photon-phonon coupling is sensitive to the stress wave amplitude. Each individual sideband transition strength is governed by a Bessel function. Considering that NV centers are point defects, sideband spectroscopy through PLE can then be used for accurate local (AC) strain/stress detection, which acts as a straightforward method for resonator characterization.

\begin{figure}[!ht]
\centering
\includegraphics{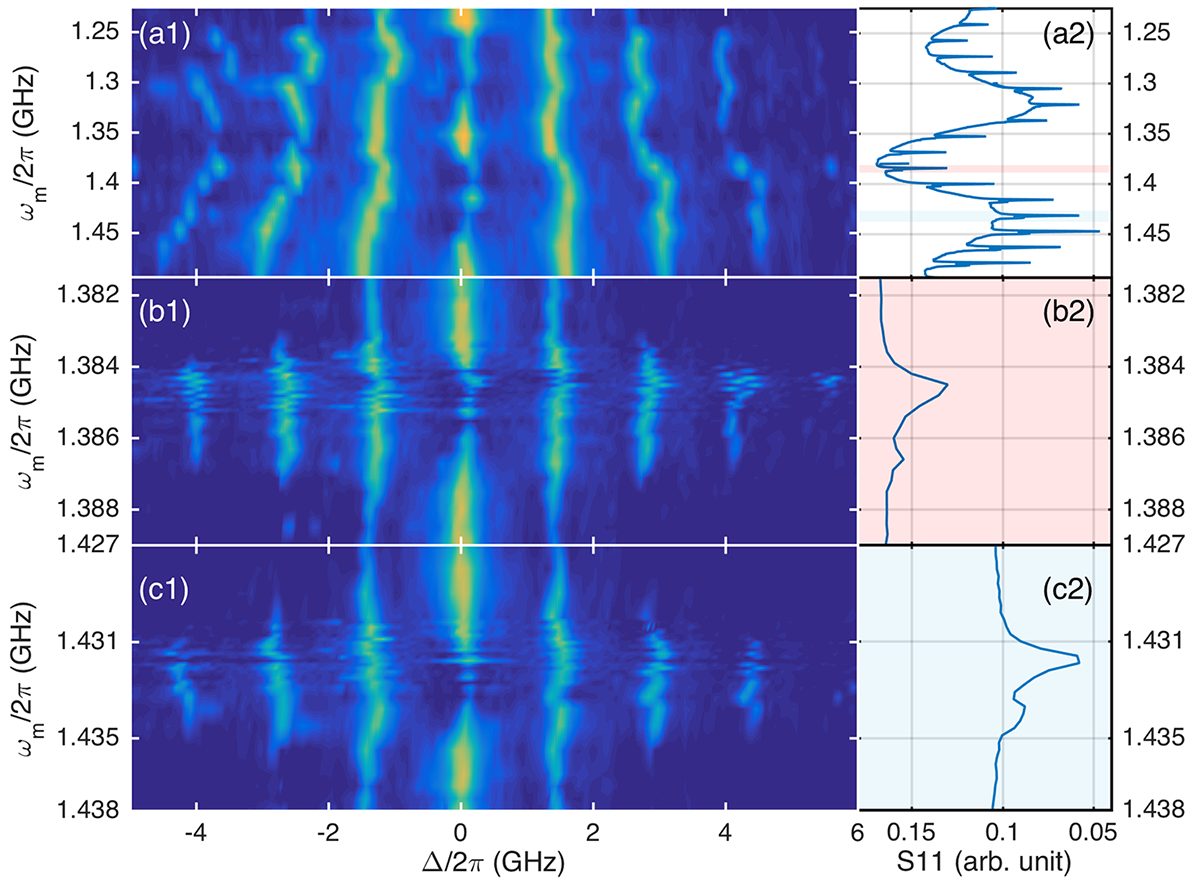}
\caption{(a) With a fixed mechanical driving power of 1 mW on the transducer, the driving frequency is scanned through the resonant mechanical modes of the HBAR device which features resonator modes spanning a range from 1.0 to 1.6 GHz, separated by a free spectral range of 16.7 MHz. The PLE spectra is shown in (a1) and the corresponding reflected power as a function of driving frequency, S11, is plotted in (a2), where a single narrow dip corresponds to a mechanical resonance mode. The wiggles in the PLE spectra are due to the change in the relative position of the NV to the anti-node of the stress wave with $\omega_{m}$ and spectral diffusion.  In (b) and (c), a close frequency scan around two individual modes, $\omega_{m}/2\pi$= 1.3844 GHz and 1.4315 GHz. The number of observable sidebands in the NV center PLE spectra is the largest when $\omega_{m}$ is close to the HBAR resonance frequency.}
\label{fig:resonator}
\end{figure}

In Fig.~\ref{fig:resonator}, we show PLE spectra of a single NV center $\ket{^{3}A_{2}, m_{s}=0} \rightarrow \ket{E_{x}}$  transition at a fixed mechanical driving power of 1 mW while the driving frequency is swept across resonances of the resonator. The resonance features are accurately captured in the NV sideband distribution.

\subsection{Mechanical Rabi flopping of orbital states through resonant phonon driving}

\begin{figure}[!ht]
\centering
\includegraphics{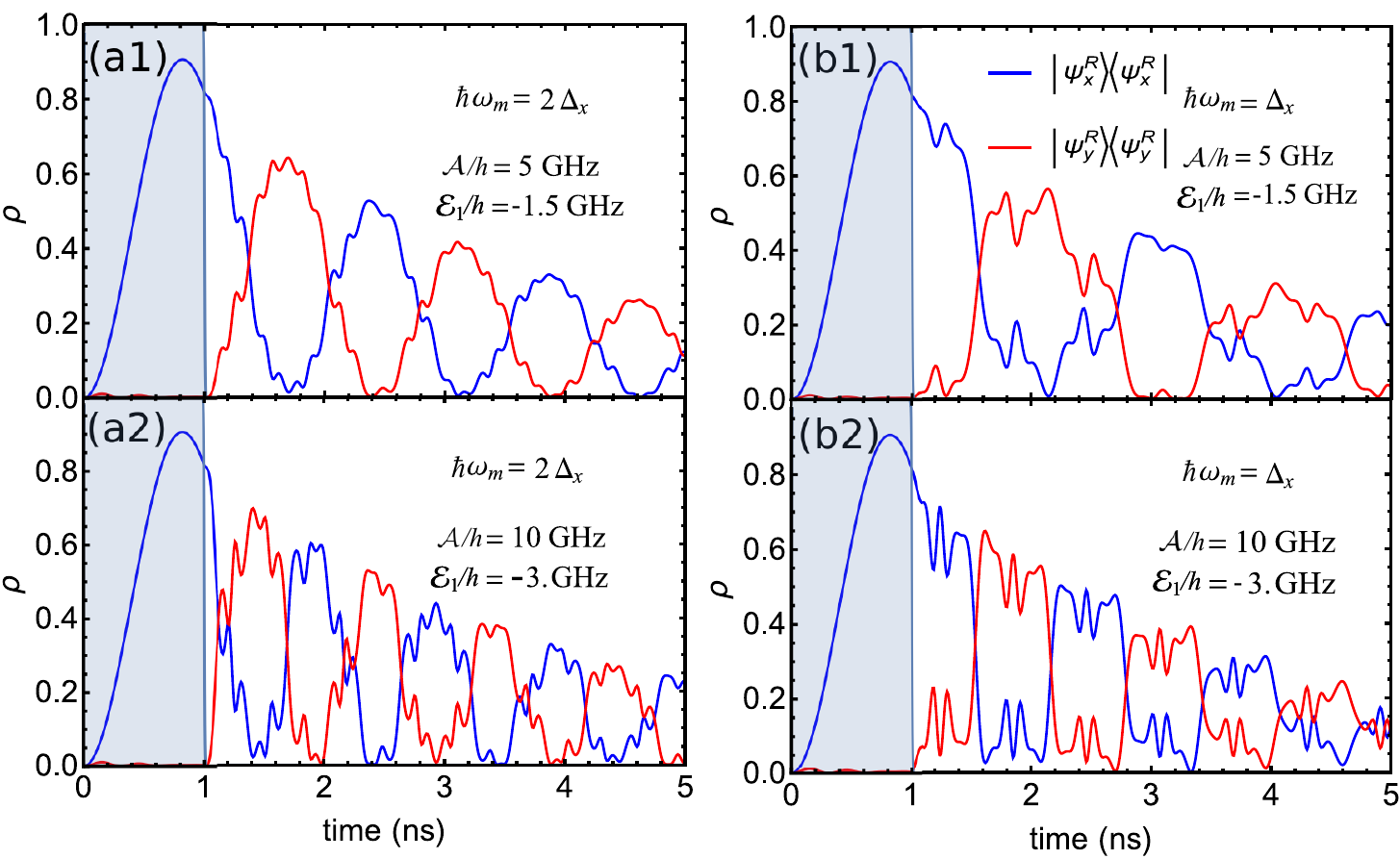}
\caption{ Phonon-driven orbital Rabi Oscillations. In each simulation, a $\pi$/2 resonant optical pulse (blue shaded region, t$<$1 ns) excites the NV from  $A\rightarrow E_{x}$, and the ensuing mechanical driving(t $>$1 ns) induces Rabi Flopping between $E_{x}$ and $E_{y}$ orbitals. The occupation probability of $E_{x}$($E_{y}$) orbital state is plotted in blue(red) curve. The sudden jumps in the simulation correspond to Landau-Zener transitions between orbitals~\cite{PRA2007}. (a) Single phonon transition, $\hbar\omega_{m}=2\Delta_{x}$. (b) 1+1 phonon transition, $\hbar\omega_{m}=\Delta_{x}$. The mechanical driving power in (a2) and (b2) is twice as that in (a1) and (b1). The parameters for simulation used here are referenced from Fig.~\ref{fig:phonontransition}(c): $2\Delta_{x}/\hbar=$3.24 GHz, $\Omega/2\pi$=0.5 GHz, $\Gamma/2\pi$=0.05 GHz.}
\label{fig:Rabi}
\end{figure}

Orbital-phonon interaction as we discussed in section~\ref{Ans} provides new ways to control NV center orbital states within the excited-state manifold. In Fig.~\ref{fig:Rabi}, we simulate the orbital state evolution under resonant phonon driving, 2$\Delta_{x}$=$(n+1)\hbar\omega_{m}$, for the NV center (NV2) studied in Fig.~\ref{fig:phonontransition} (c). The simulation results show single and multi-phonon transitions between $\ket{E_{x}}$ and $\ket{E_{y}}$ orbitals. Fig.~\ref{fig:Rabi}(b2) shows that at the largest mechanical driving power used in Fig.~\ref{fig:phonontransition}(d), the phonon driven orbital Rabi approaches a nanosecond.

\subsection{Continuous dynamical decoupling of NV center orbital states}

As shown in Fig.~\ref{fig:level}(b), NV center excited orbital states, $\ket{E_{x,y}}$ ($\ket{E_{1,2}}$ and $\ket{A_{1,2}}$), couple (nearly) linearly to transverse strain field perturbation. Similar dispersion relation can be found for electric fields through the Stark effect~\cite{Maze2011}. Fluctuations in the ambient electric field from, for example, charge fluctuations in the NV center's immediate environment, can result in unstable orbital states and spectral diffusion in the optical transitions. This harms the optical coherence for quantum optical control and other applications of resonant optical coupling.

\begin{figure}[!ht]
\centering
\includegraphics{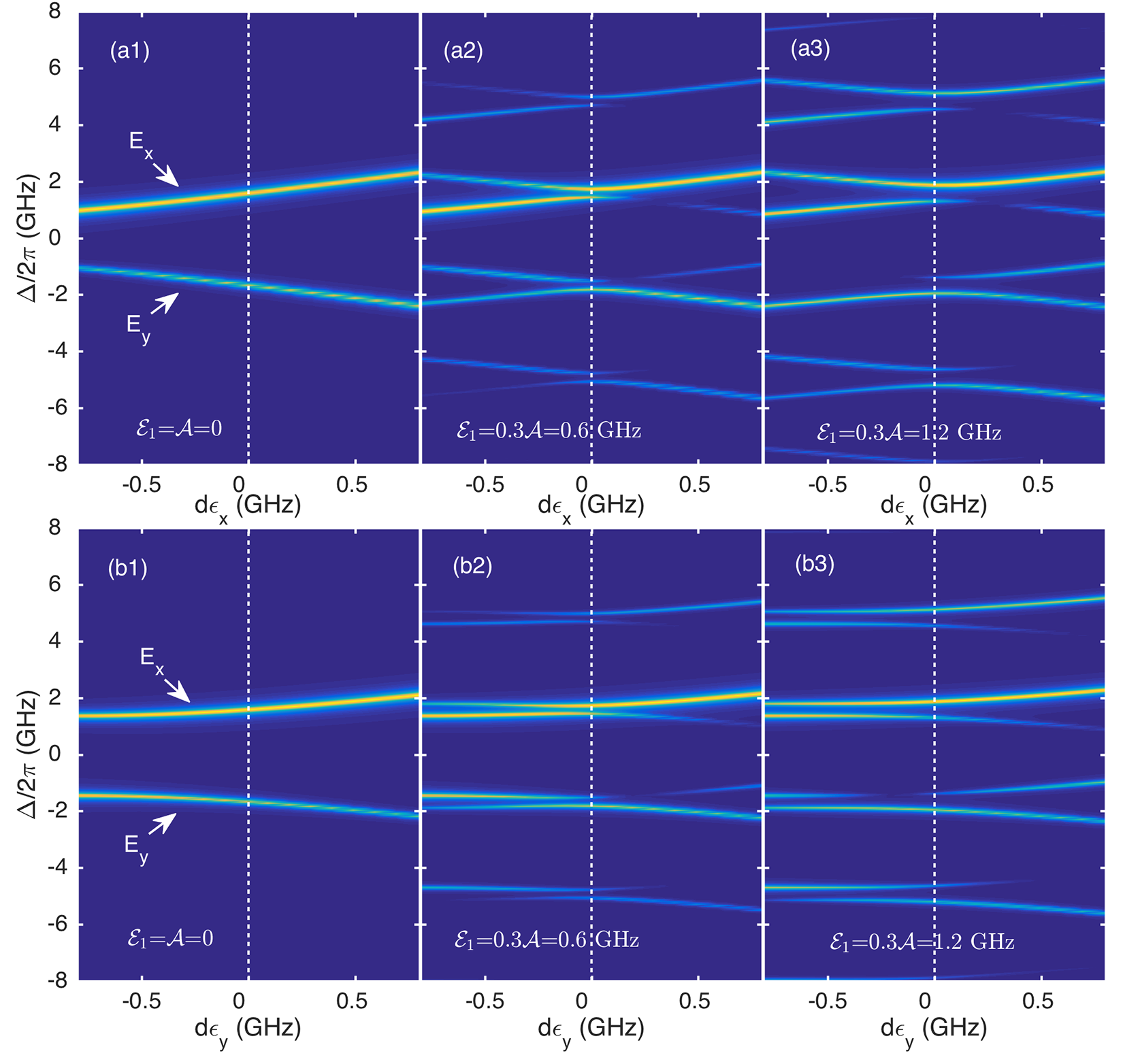}
\caption{Simulated electric field energy dispersion of NV2 $\ket{E_{x,y}}$ states under continuous resonant phonon driving, $\hbar\omega_{m}=2\Delta_{x}$, with increasing $\mathcal{E}_{1}$ modulation strength from (1-3). d$\epsilon_{x}$ (a1-a3) and d$\epsilon_{y}$ (b1-b3) are the transverse (with respective to NV axis) electric dipole interactions. As seen from the simulation (a2, b2), $d\omega/d\epsilon_{\perp}$ is close to 0 under 0.6 GHz $\mathcal{E}_{1}$ driving. }
\label{fig:CDD}
\end{figure}

\begin{figure}[!ht]
\centering
\includegraphics{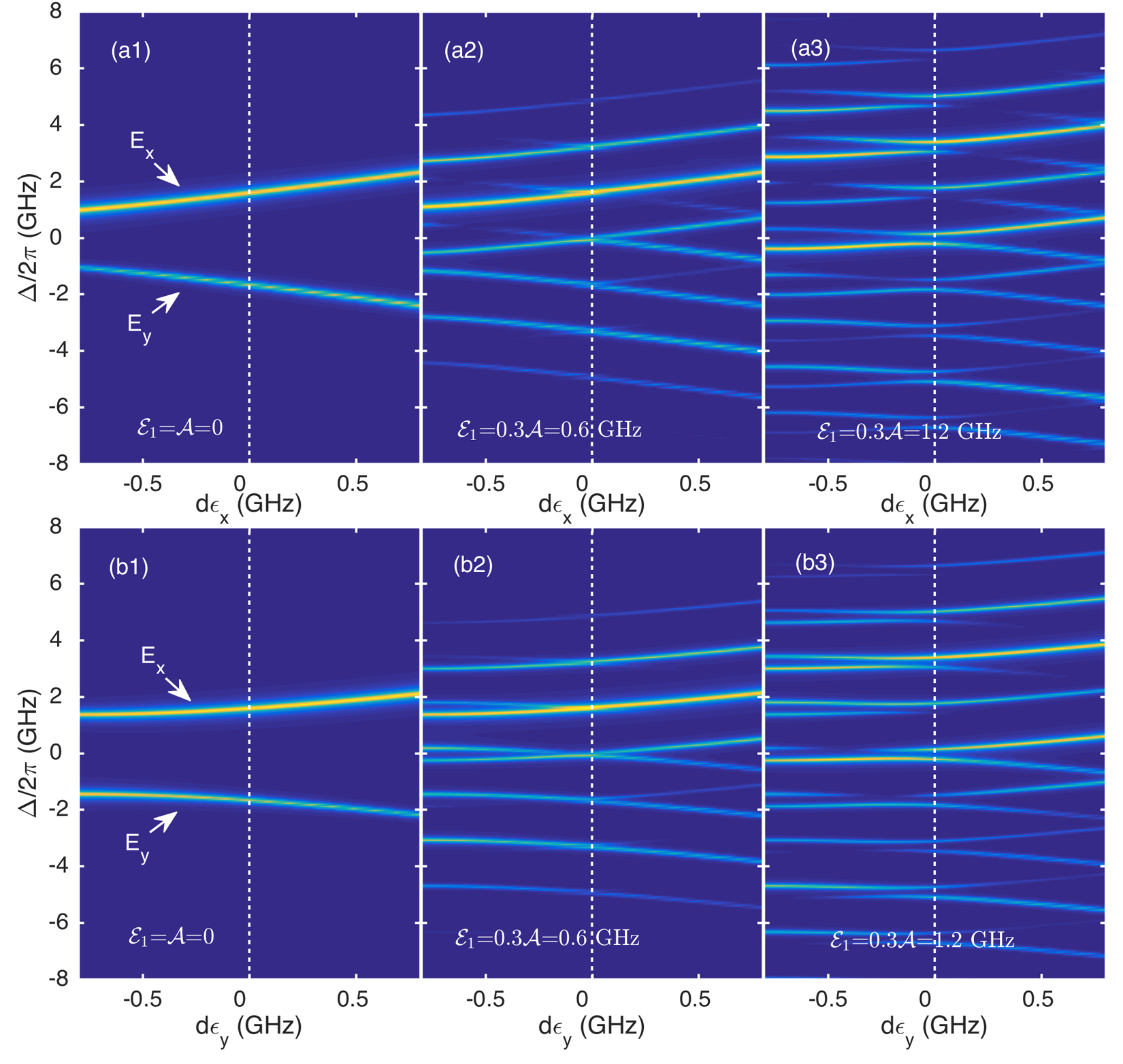}
\caption{Simulated electric field energy dispersion of NV2 $\ket{E_{x,y}}$ $\ket{E_{x,y}}$ states under continuous resonant phonon driving, $\hbar\omega_{m}=\Delta_{x}$, with increasing $\mathcal{E}_{1}$ modulation strength from (1-3). As seen from the simulation (a2, b2), $d\omega/d\epsilon_{\perp}$ is close to 0 under 1.2 GHz $\mathcal{E}_{1}$ driving.}
\label{fig:CDD2}
\end{figure}

Phonon-dressed NV center orbital states have a different energy dispersion relation than that of the unperturbed orbital states. In certain conditions, for example under continuous resonant orbital driving, they can be engineered to become less sensitive to electric field fluctuations. This general approach is known as continuous dynamical decoupling (CDD)~\cite{Fanchini07}, which has not, to our knowledge, previously been applied to orbital states. In Fig.~\ref{fig:CDD} and Fig.~\ref{fig:CDD2}, we simulate the energy dispersion of $\ket{E_{x}}$ and $\ket{E_{y}}$ states of NV2 subject to resonant phonon driving with $\hbar\omega_{m}=2\Delta_{x}$ and $\hbar\omega_{m}=\Delta_{x}$, respectively. The phonon-dressed orbital states feature $d\omega/d\epsilon_{\perp}$ close to 0 and are less sensitive to electric field fluctuations.

\clearpage
\bibliographystyle{apsrev4-1}
\bibliography{NVPLEpaper}

\end{document}